\documentclass[aps,prd,twocolumn,10pt,superscriptaddress,nofootinbib,nobibnotes,longbibliography]{revtex4-1}

\usepackage{amssymb}
\usepackage{graphicx}
\usepackage{amsmath}
\usepackage{hyperref}
\usepackage{subfigure}
\usepackage{multirow}
\usepackage{setspace}
\usepackage{verbatim}
\usepackage{float}
\usepackage{color}
\usepackage{ulem}
\usepackage[utf8]{inputenc}
\usepackage[table,xcdraw]{xcolor}
\usepackage{makecell}

\begin{document}

\title{General backreaction force of cosmological bubble expansion}

\author{Jun-Chen Wang}
\email{junchenwang@stu.pku.edu.cn}
\affiliation{School of Physics, Peking University, Beijing 100871, China}

\author{Zi-Yan Yuwen}
\email{yuwenziyan@itp.ac.cn (Corresponding author)}
\affiliation{CAS Key Laboratory of Theoretical Physics, Institute of Theoretical Physics, Chinese Academy of Sciences (CAS), Beijing 100190, China}
\affiliation{School of Physical Sciences, University of Chinese Academy of Sciences (UCAS), Beijing 100049, China}

\author{Yu-Shi Hao}
\email{haoyushi@itp.ac.cn}
\affiliation{CAS Key Laboratory of Theoretical Physics, Institute of Theoretical Physics, Chinese Academy of Sciences (CAS), Beijing 100190, China}
\affiliation{School of Physical Sciences, University of Chinese Academy of Sciences (UCAS), Beijing 100049, China}

\author{Shao-Jiang Wang}
\email{schwang@itp.ac.cn (Corresponding author)}
\affiliation{CAS Key Laboratory of Theoretical Physics, Institute of Theoretical Physics, Chinese Academy of Sciences (CAS), Beijing 100190, China}
\affiliation{Asia Pacific Center for Theoretical Physics (APCTP), Pohang 37673, Korea}

%\date{\today}

\begin{abstract}
The gravitational-wave energy-density spectra from cosmological first-order phase transitions crucially depend on the terminal wall velocity of asymptotic bubble expansion when the driving force from the effective potential difference is gradually balanced by the backreaction force from the thermal plasma. Much attention has previously focused on the backreaction force acting on the bubble wall alone but overlooked the backreaction forces on the sound shell and shock-wave front, if any, which have been both numerically and analytically accomplished in our previous studies but only for a bag equation of state. In this paper, we will generalize the backreaction force on bubble expansion beyond the simple bag model.
\end{abstract}
\maketitle

\section{Introduction}\label{sec:introduction}

For new physics beyond the standard model of particle physics by breaking some continuous symmetry~\cite{Caldwell:2022qsj}, our early universe would experience a cosmological first-order phase transition (FOPT)~\cite{Mazumdar:2018dfl,Hindmarsh:2020hop} if there is a potential barrier separating the false and true vacua with decreasing temperature~\cite{Quiros:1999jp}. This quantum phase transition proceeds dominantly via stochastic nucleations of true vacuum bubbles as a Poisson process~\cite{Athron:2023xlk}, which also universally induces curvature perturbations~\cite{Liu:2022lvz} or even primordial black holes~\cite{Liu:2021svg,Hashino:2021qoq,He:2022amv} due to the anisochronous nature of bubble nucleations (see also Refs.~\cite{Lewicki:2023ioy,Gouttenoire:2023naa,Baldes:2023rqv} for recent improvements). After bubble nucleations, the expansion of bubble walls drags and/or pushes the thermal fluids of background plasma to form surrounding sound shells, whose collisions (and subsequent evolution into magnetohydrodynamic turbulence) would generate sound waves as the dominant stochastic gravitational wave background (SGWB) over that from bubble-wall collisions~\cite{Weir:2017wfa}. The SGWB from bubble-wall collisions can also dominate over that from sound waves when the backreaction from thermal plasma against expansion grows rather slowly or the initial average separation of bubbles is quite close so that the accelerating walls never had a chance to actually approach the terminal velocity before they have already largely collided with each other~\cite{Cai:2020djd,Lewicki:2022pdb}. Future detection of these SGWBs from FOPTs at different frequency bands~\cite{Caprini:2015zlo,Caprini:2019egz} would reveal new physics~\cite{Cai:2017cbj,Bian:2021ini} beyond the standard model of particle physics at different temperature scales.

However, the energy-density spectra of SGWBs from FOPTs have not been fully determined yet. For example, recent numerical simulation~\cite{Sharma:2023mao} and analytical estimation~\cite{RoperPol:2023dzg} of sound waves seem to recognize a steep growth with $k^9$-scaling near the peak instead of previously thought in the far infrared~\cite{Hindmarsh:2019phv,Hindmarsh:2016lnk} since, and the infrared regime should preserve the causal $k^3$-scaling~\cite{Cai:2019cdl} as shown recently in a hydrodynamic sound-shell model~\cite{Cai:2023guc}. Despite the successes of characterizing the GW spectra from FOPTs by a handful of phenomenological parameters, we highlight here two aspects of complexities that make the model predictions rather difficult to be precise in order to be tested efficiently against future observational data. The first complexity comes from the efficiency factor $\kappa_v$ of converting released vacuum energy into bulk fluid motions. Although it can be well-fitted analytically for a bag equation of state (EOS)~\cite{Espinosa:2010hh} regardless of the details of underlying particle physics, the $\nu$-model predictions~\cite{Giese:2020znk,Giese:2020rtr,Wang:2020nzm,Wang:2023jto} beyond the simple bag EOS become more involved with the particle physics, which is even worse for the case beyond the $\nu$-model when further considering the varying sound velocity effect~\cite{Wang:2022lyd} in the sound shell. The second complexity comes from the terminal wall velocity that is closely related to the out-of-equilibrium effect~\cite{Moore:1995ua,Moore:1995si,Konstandin:2014zta,Laurent:2020gpg,Dorsch:2021nje,Laurent:2022jrs} in the vicinity of the bubble wall.

The terminal wall velocity can be determined in principle if the backreaction force can be known exactly~\cite{Ellis:2019oqb,Ellis:2020nnr,Cai:2020djd,Lewicki:2022pdb}. Therefore, much attention has previously focused on calculating the pressure recoil\footnote{This is what is usually called the friction force. However, in our previous studies~\cite{Wang:2022txy,Li:2023xto} and the current paper, we have a more rigorous definition for the friction force as seen below.} from interacting particles acting on the wall alone but usually in the ultrarelativistic limit~\cite{Bodeker:2009qy,Bodeker:2017cim,BarrosoMancha:2020fay,Hoche:2020ysm,Azatov:2020ufh,Gouttenoire:2021kjv,DeCurtis:2022hlx}. It was later realized in Ref.~\cite{Ai:2021kak} that the so-called friction force from the nonequilibrium effect is not the whole story but the inhomogeneous temperature profile also contributes as a thermal force to the total backreaction force~\cite{Wang:2022txy,Li:2023xto}.\footnote{Recall that some earlier studies~\cite{Espinosa:2010hh,Konstandin:2010dm} have attributed this inhomogeneous temperature contribution to a modification of the driving force. Here we take a different but more physical viewpoint as elaborated below.} Without assuming a bag EOS and without requiring an equilibrium distribution function, we have proposed in Ref.~\cite{Wang:2022txy} the exact hydrodynamic formula for calculating the total backreaction force including the thermal force and friction force for a nonrunaway steady-state bubble expansion, which is verified numerically~\cite{Wang:2022txy} and proved analytically~\cite{Li:2023xto} for the simple case with a bag EOS. In particular, the hydrodynamic computation for full backreaction force can be decomposed into the part that acts on the sound shell and the part that acts on the discontinuous interface (such as the bubble wall and shock-wave front, if any), the latter of which is shown to be a highly nontrivial contribution in order to balance the driving force from the effective-potential difference. The take-home message from our studies is that it is actually not enough to fully determine the wall velocity simply from the pressure recoil on the wall alone, and we must know all about the full backreaction force acting on the bubble wall, sound shell, and shock-wave front, if any.

In this paper, we will generalize our previous studies~\cite{Wang:2022txy,Li:2023xto} into the $\nu$-model beyond the simple bag EOS when the sound velocities are two different constants inside and outside of the bubble wall. We will first set up our conventions and review our previous results in Sec.~\ref{sec:setup}, and then give rise to the general backreactioin force beyond the bag EOS in Sec.~\ref{sec:pbrmuModel}, which can be analytically proved in Sec.~\ref{sec:ProofBalance} from two consistency checks. We finally conclude in Sec.~\ref{sec:conclusion} with a concise summary of our analytic results and then discuss in Sec.~\ref{sec:discussions} several applications and generalizations of our hydrodynamic approach in future works.

\section{General backreaction of bubble expansion}\label{sec:setup}

In this section, we will briefly review and slightly improve our previous results in Refs.~\cite{Wang:2022txy,Li:2023xto}, including the equation of motions from the scalar-plasma system, the energy-momentum tensor of the approximated wall-fluid system, and in particular the general form of the backreaction force from hydrodynamics.

\subsection{The scalar-plasma system}\label{subsec:scalarplasma}

The cosmological FOPT as a typical scalar-plasma system admits its total energy-momentum tensor $T^{\mu\nu}=T_\phi^{\mu\nu}+T_f^{\mu\nu}$ by simply adding up the scalar field and thermal plasma parts,
\begin{align}
T_\phi^{\mu\nu}&=\nabla^\mu\phi\nabla^\nu\phi+g^{\mu\nu}\left[-\frac12(\nabla\phi)^2-V_0(\phi)\right],\label{eq:Tphi}\\
T_f^{\mu\nu}&=\sum\limits_{i=\mathrm{B,F}}g_i\int\frac{\mathrm{d}^3\mathbf{k}}{(2\pi)^3}\frac{k^\mu k^\nu}{k^0}\bigg|_{k^0=E_i(\mathbf{k})}f_i(\mathbf{x},\mathbf{k}),\label{eq:Tpla}
\end{align}
where $f_i(\mathbf{x},\mathbf{k})$ is the distribution function counting the average number of particles of species $i$ with momentum $\mathbf{k}$, effective mass $m_i(\phi)$, and energy $E_i(\mathbf{k})\equiv\sqrt{\mathbf{k}^2+m_i^2}$ in a volume element $(\mathbf{x},\mathbf{x}+\mathrm{d}\mathbf{x})\times(\mathbf{k},\mathbf{k}+\mathrm{d}\mathbf{k})$ of the phase space at the time $t=x^0$. These distribution functions are governed by their own relativistic Boltzmann equations,
\begin{align}
\frac{\mathrm{D}}{\mathrm{d}\lambda}f_i(x^\mu,k^\mu)=\left(\frac{\mathrm{D}x^\mu}{\mathrm{d}\lambda}\nabla_\mu+\frac{\mathrm{D}k^\mu}{\mathrm{d}\lambda}\frac{\partial}{\partial k^\mu}\right)f_i=C[f_i]
\end{align}
where the affine parameter for a massive particle of mass $m_i$ takes $\lambda=\tau/m_i$ with $\mathrm{d}\tau^2=-\mathrm{d}s^2=-g_{\mu\nu}\mathrm{d}x^\mu\mathrm{d}x^\nu$, while for a massless particle, the affine parameter takes the form that exactly defines its momentum four-vector $k^\mu=\mathrm{d}x^\mu/\mathrm{d}\lambda$ along its null geodesic $x^\mu(\lambda)$. $C[f_i]$ is the collision term to species $i$. Here the two directional covariant derivative terms simply define the four momentum and external force,
\begin{align}
\frac{\mathrm{D}x^\mu}{\mathrm{d}\lambda}=\frac{\mathrm{d}x^\mu}{\mathrm{d}\lambda}=k^\mu,\quad \frac{\mathrm{D}k^\mu}{\mathrm{d}\lambda}=\frac{\mathrm{d}k^\mu}{\mathrm{d}\lambda}+\Gamma_{\nu\sigma}^\mu k^\nu k^\sigma=m_i F_i^\mu.
\end{align}
For the worldline action $S=-\int m_i\mathrm{d}\tau\,\sqrt{-g_{\mu\nu}\mathrm{d}\dot{x}^\mu\mathrm{d}\dot{x}^\nu}$ with $\dot{x}^\mu\equiv\mathrm{d}x^\mu/\mathrm{d}\tau$ of a single particle moving with a field-dependent mass $m_i(\phi)$ in the background scalar field $\phi(x^\mu)$, the external force can be determined from its equation of motion (EOM) as $F_i^\mu=-m'_i(\phi)\nabla^\mu\phi$. Therefore, the final relativistic Boltzmann equation for each particle species $i$ on shell reads
\begin{align}
\left(k^\mu\nabla_\mu-\frac12\nabla^\mu\phi\frac{\mathrm{d}m_i^2}{\mathrm{d}\phi}\frac{\partial}{\partial k^\mu}\right)\Theta(k^0)\delta(k^2+m_i^2)f_i(x,k)=C[f_i].
\end{align}

The dynamics of the scalar-plasma system for the cosmological FOPT is governed by the conservation of the total energy-momentum tensor $\nabla_\mu(T_\phi^{\mu\nu}+T_f^{\mu\nu})=0$ but with each term broken separately as
\begin{align}
\nabla_\mu T_\phi^{\mu\nu}&\equiv[\nabla_\mu\nabla^\mu\phi-V'_0(\phi)]\nabla^\nu\phi=+f^\nu,\\
\nabla_\mu T_f^{\mu\nu}&\equiv\sum\limits_{i=\mathrm{B,F}}g_i\int\frac{\mathrm{d}^3\mathbf{k}}{(2\pi)^3}\frac{k^\mu k^\nu}{E_i(\mathbf{k})}\nabla_\mu f_i=-f^\nu
\end{align}
by the dubbed transfer flow $f^\nu$~\cite{Moore:1995ua,Moore:1995si} (see also~\cite{Hindmarsh:2020hop}),
\begin{align}\label{eq:TransferFlow}
f^\nu=\nabla^\nu\phi\sum\limits_{i=\mathrm{B,F}}g_i\frac{\mathrm{d}m_i^2}{\mathrm{d}\phi}\int\frac{\mathrm{d}^3\mathbf{k}}{(2\pi)^3}\frac{f_i}{2E_i}.
\end{align}
In particular, for the thermal equilibrium distribution functions of the Bose-Einstein/Fermi-Dirac distributions with negligible chemical potentials,
\begin{align}
f_i^\mathrm{eq}(\mathbf{x},\mathbf{k})=\frac{1}{e^{E_i(\mathbf{k})/T}\mp1},
\end{align}
the corresponding transfer flow reproduces exactly the field derivative of the one-loop part $\Delta V_T^{(1)}(\phi,T)$ of the thermal correction $\Delta V_T(\phi,T)=\Delta V_T^{(1)}(\phi,T)+ \Delta V_T^{(>1)}(\phi,T)$ to the total effective potential $V_\mathrm{eff}(\phi,T)=V_0(\phi)+\Delta V_T(\phi,T)$,
\begin{align}
\frac{\partial\Delta V_T^{(1)}}{\partial \phi}
&=\frac{\partial}{\partial\phi}\sum_{i=\mathrm{B,F}}\pm g_i T\int\frac{\mathrm{d}^3\mathbf{k}}{(2\pi)^3}\log\left(1\mp e^{-\frac{E_i(\mathbf{k})}{T}}\right),\nonumber\\
&=\sum\limits_{i=\mathrm{B,F}}g_i\frac{\mathrm{d}m_i^2}{\mathrm{d}\phi}\int\frac{\mathrm{d}^3\mathbf{k}}{(2\pi)^3}\frac{f_i^\mathrm{eq}}{2E_i}.
\end{align}

This similarity inspires us to parametrize the transfer flow in such a way that there is a one-to-one correspondence to each part of the effective potential,
\begin{align}\label{eq:TransferFlowAnsatz}
f^\nu=\nabla^\nu\phi\left(\frac{\partial \Delta V_T^{(1)}}{\partial \phi}+\frac{\partial \Delta V_T^{(>1)}}{\partial \phi}-\frac{\partial p_{\delta f}}{\partial\phi}\right),
\end{align}
where the first, second, and third terms correspond to the equilibrium distribution function $f_i^\mathrm{eq}$ of free particles, the equilibrium distribution function $\Delta f_i^\mathrm{eq}$ with interacting particles, and the nonequilibrium part $\delta f_i$ of the total distribution function $f_i=f_i^\mathrm{eq}+\Delta f_i^\mathrm{eq}+\delta f_i$, respectively. With the above ansatz for the transfer flow, the conservation equations of the scalar field and thermal plasma can now be rearranged as
\begin{align}
\nabla_\mu\nabla^\mu\phi-\frac{\partial V_\mathrm{eff}}{\partial\phi}&=-\frac{\partial p_{\delta f}}{\partial\phi},\label{eq:EOMscalar}\\
\nabla_\mu T_f^{\mu\nu}+\nabla^\nu\phi\frac{\partial \Delta V_T}{\partial\phi}&=\nabla^\nu\phi\frac{\partial p_{\delta f}}{\partial\phi},\label{eq:EOMplasma}
\end{align}
as also expected from the Kadanoff-Baym equations~\cite{Konstandin:2014zta}. Note that our ansatz is different from the usual one
\begin{align}
f^\nu=\nabla^\nu\phi\left(\frac{\partial \Delta V_T^{(1)}}{\partial \phi}-\frac{\partial p_{\delta f}}{\partial\phi}\right),
\end{align}
by just splitting the total distribution function $f_i=f_i^\mathrm{eq}+\delta f_i$ into the equilibrium and nonequilibrium parts, in which case the conservation equations of the scalar field and thermal plasma now read
\begin{align}
\nabla_\mu\nabla^\mu\phi-\frac{\partial V_\mathrm{eff}^{(1)}}{\partial\phi}&=-\frac{\partial p_{\delta f}}{\partial\phi},\\
\nabla_\mu T_f^{\mu\nu}+\nabla^\nu\phi\frac{\partial \Delta V_T^{(1)}}{\partial\phi}&=\nabla^\nu\phi\frac{\partial p_{\delta f}}{\partial\phi},
\end{align}
with the effective potential $V_\mathrm{eff}^{(1)}$ and thermal correction $\Delta V_T^{(1)}$ only up to the one-loop order. We will leave this ambiguity for future study. Fortunately, this ambiguity would not affect our hydrodynamic determination of the general backreaction force.

\subsection{The wall-fluid system}\label{subsec:wallfluid}

The general EOMs \eqref{eq:EOMscalar} and \eqref{eq:EOMplasma} of the scalar-plasma system for the cosmological FOPT are difficult to solve as the out-of-equilibrium term $\partial p_{\delta f}/\partial\phi$ cannot be known in general due to the highly model-dependent collision term $C[f_i]$. To derive our general backreaction force, we further assume two major simplifications to approximate the scalar-plasma system into the dubbed wall-fluid system.

The first simplification is to consider a fast FOPT so that it can be completed within one Hubble time and hence the background spacetime can be approximated as a flat spacetime, $g_{\mu\nu}=\eta_{\mu\nu}$. In the flat spacetime, we can fix the bubble nucleation site at the origin point, and then build different coordinate systems depending on geometries of the bubble wall. For bubble expansion of planar, cylindrical, and spherical walls, we can choose corresponding systems with $x^\mu=(t,z,x=0,y=0)$, $x^\mu=(t,\rho,\varphi=0,z=0)$, and $x^\mu=(t,r,\theta=0,\varphi=0)$, respectively. After setting up the coordinate system, we can define the plasma rest frame comoving with all particle species (assuming moving coherently\footnote{Otherwise, it could be a two-fluid or even multifluid system, which can be more difficult and is reserved for future study.}) with four-velocity $u^\mu=(1,0,0,0)$. In this plasma rest frame, we can further define the plasma energy density $e_f$ and pressure $p_f$ (only along the radial $x^1$ direction\footnote{Otherwise, the plasma species can move along $x^2$ and $x^3$ directions with shear and viscosity, which can be more difficult and is reserved for future study.}) locally as
\begin{align}
e_f&=\sum\limits_{i=\mathrm{B,F}}g_i\int\frac{\mathrm{d}^3\mathbf{k}}{(2\pi)^3}E_i(\mathbf{k})f_i,\\
p_f&=\sum\limits_{i=\mathrm{B,F}}g_i\int\frac{\mathrm{d}^3\mathbf{k}}{(2\pi)^3}\frac{\mathbf{k}^2}{E_i(\mathbf{k})}f_i,
\end{align}
with which the energy-momentum tensor~\eqref{eq:Tpla} of the thermal plasma can be rearranged into a perfect-fluid form,
\begin{align}\label{eq:plasmafluid}
T_f^{\mu\nu}
&=\sum\limits_{i=\mathrm{B,F}}g_i\int\frac{\mathrm{d}^3\mathbf{k}}{(2\pi)^3}\left[\left(E_i+\frac{\mathbf{k}^2}{E_i}\right)u^\mu u^\nu+\eta^{\mu\nu}\frac{\mathbf{k}^2}{E_i}\right]f_i\nonumber\\
&\equiv(e_f+p_f)u^\mu u^\nu+p_f\eta^{\mu\nu}.
\end{align}

The second simplification is to consider the late stage of the fast FOPT, which further leads to three natural consequences: (1) self-similar expansion; (2) thin-wall expansion; and (3) steady-state expansion. When considering the late stage, the initial bubble size is negligible, and hence, there is no more characteristic scale; thus, the bubble expansion is self-similar and characterized by a single self-similar coordinate $\xi=x^1/x^0$. At the late stage of bubble expansion, the bubble wall can be approximated as a thin wall since the wall thickness actually decreases with the expanding wall radius. For a nonrunaway expansion as expected from recent debate~\cite{Bodeker:2009qy,Bodeker:2017cim,Hoche:2020ysm,Gouttenoire:2021kjv}, the late-stage expansion would reach a steady state with a terminal wall velocity denoted as $\xi_w$ hereafter.

With all the above three approximations, the scalar field profile along the $x^1$ direction can be written as  $\phi(x^0,x^1)=\phi_+\Theta(x^1/x^0-\xi_w)+\phi_-\Theta(\xi_w-x^1/x^0)\equiv\phi(x^1/x^0\equiv\xi)$ interpolating the false vacuum $\phi_+$ and true vacuum $\phi_-$ with a steplike function so that the derivative $\phi'(\xi)=(\phi_+-\phi_-)\delta(\xi-\xi_w)$ exhibits a Dirac-delta singularity. As we will see later, it is this Dirac-delta behavior of the scalar wall and thermal plasma at the bubble wall and shock-wave front (if any) that admits nontrivial contributions to the backreaction force. With the above explicit scalar profile, we can directly compute the energy-momentum tensor first in the wall frame and then back to the plasma frame. In a local frame comoving with the bubble wall, the local wall geometry can be approximated as a planar wall, and hence, a Lorentz transformation $\bar{t}\equiv\gamma_w(t-\xi_wz)$ and $\bar{z}\equiv\gamma_w(z-\xi_wt)$ can be defined between the wall frame $(\bar{t},\bar{z})$ and plasma frame $(t,z)$ with $\gamma_w=(1-\xi_w)^{-1/2}$ the Lorentz factor of terminal wall velocity. The benefit of going to the wall frame is that we can remove the time derivative term $\partial_{\bar{t}}\phi=(\gamma_w/t)(\xi_w-\xi)\phi'(\xi)=(\gamma_w/t)(\xi_w-\xi)(\phi_+-\phi_-)\delta(\xi-\xi_w)=0$, and hence, the energy-momentum tensor of the scalar wall contains nonvanishing components only from $T_\phi^{\bar{t}\bar{t}}= V_0$ and $T_\phi^{\bar{z}\bar{z}}=-V_0$. When transformed back to the background plasma frame with the Lorentz transformations  $T_\phi^{\bar{t}\bar{t}}=\gamma_w^2(T_\phi^{tt}+T_\phi^{zz})-T_\phi^{zz}$ and $T_\phi^{\bar{z}\bar{z}}=(\gamma_w^2-1)(T_\phi^{tt}+T_\phi^{zz})+T_\phi^{zz}$, we can solve for $T_\phi^{tt}=T_\phi^{\bar{t}\bar{t}}=V_0\equiv e_\phi$ and $T_\phi^{zz}=T_\phi^{\bar{z}\bar{z}}=-V_0\equiv p_\phi$ that can be easily rearranged into a perfect-fluid $T_\phi^{\mu\nu}=(e_\phi+p_\phi)u^\nu u^\nu+p_\phi\eta^{\mu\nu}$ with vanishing enthalpy $w_\phi=e_\phi+p_\phi=0$.

In a short summary, both energy-momentum tensors from the scalar wall and thermal plasma can be cast into a perfect-fluid form; therefore, they can be naively combined into a total energy-momentum tensor of the same form $T^{\mu\nu}=(e+p)u^\mu u^\nu+p\eta^{\mu\nu}$ with $e=e_f+e_\phi=e_f+V_0$ and $p=p_f+p_\phi=p_f-V_0$. An immediate observation $p_f=p-p_\phi=-V_\mathrm{eff}+V_0=-\Delta V_T$ can be made from the definition $-p=\mathcal{F}=V_\mathrm{eff}=V_0+\Delta V_T$. By introducing an abbreviation $\mu(\zeta, v(\xi))\equiv(\zeta-v)/(1-\zeta v)$ for the Lorentz transformation of the bulk fluid velocity $v(\xi)$ in the plasma frame into a local frame comoving with a velocity $\zeta$, we can define the wall-frame fluid velocity $\bar{v}(\xi)=\mu(\xi_w,v(\xi))$ with corresponding Lorentz factor $\bar{\gamma}=(1-\bar{v})^{-1/2}$. Hence, the four-velocity reads $u^\mu=\bar{\gamma}(1,-\bar{v},0,0)$ in the local wall frame, where the minus sign is introduced to ensure a positive $\bar{v}$. Therefore, the total energy-momentum tensor in a local wall frame reads
\begin{align}
T^{\mu\nu}=\left(\begin{array}{cccc}
w\bar{\gamma}^2-p & -w\bar{\gamma}^2\bar{v}   & 0 & 0 \\
 -w\bar{\gamma}^2\bar{v} & w\bar{\gamma}^2\bar{v}^2+p & 0 & 0 \\
 0 & 0 & p & 0\\
 0 & 0 & 0 & p
\end{array}\right), \quad \mu,\nu=\bar{t},\bar{z},x,y.
\end{align}
A similar form can also be obtained for the total energy-momentum tensor in a local shock-wave-front frame by replacing all overbar symbols with overtilde symbols, where the shock-frame fluid velocity is defined by $\tilde{v}(\xi)=\mu(\xi_{sh},v(\xi))$ with $\xi_{sh}$ the velocity for the shock-wave front, if any.

With the perfect-fluid ansatz for the total energy-momentum tensor of the wall-fluid system, one can derive hydrodynamic EOMs from the projected conservation equations $u_\nu\nabla_\mu T^{\mu\nu}=0$ and $\tilde{u}^\nu\nabla^\mu T_{\mu\nu}=0$ parallel along and perpendicular to the fluid flow directions defined by $u^\mu=\gamma(v)(1,v,0,0)$ and $\tilde{u}^\mu=\gamma(v)(v,1,0,0)$ with $u_\mu u^\mu=-1$, $\tilde{u}_\mu\tilde{u}^\mu=1$, $u_\nu\nabla_\mu u^\nu=0$, $\tilde{u}_\mu u^\mu=0$,
\begin{align}
u^\mu\nabla_\mu e&=-w\nabla_\mu u^\mu,\\
\tilde{u}^\mu\nabla_\mu p&=-w\tilde{u}^\nu u^\mu\nabla_\mu u_\nu.
\end{align}
For bubble expansion of planar, cylindrical, and spherical wall geometries with $D=0,1,2$, respectively, we can express $\nabla_\mu u^\mu=(Dv/\xi)(\gamma/t)+(\gamma^3/t)(1-\xi v)\partial_\xi v$ explicitly in the plasma frame with the self-similar coordinate, and the projected conservation equations become
\begin{align}
(\xi-v)\frac{\partial_\xi e}{w}&=D\frac{v}{\xi}+\gamma^2(1-\xi v)\partial_\xi v,\label{eq:de}\\
(1-\xi v)\frac{\partial_\xi p}{w}&=\gamma^2(\xi-v)\partial_\xi v,\label{eq:dp}
\end{align}
which can be rearranged by division and summation as
\begin{align}
D\frac{v}{\xi}&=\gamma^2(1-\xi v)\left(\frac{\mu(\xi,v)^2}{c_s^2}-1\right)\frac{\mathrm{d}v}{\mathrm{d}\xi},\label{eq:dv}\\
\frac{\mathrm{d}w}{\mathrm{d}\xi}&=w\gamma^2\mu(\xi,v)\left(\frac{1}{c_s^2}+1\right)\frac{\mathrm{d}v}{\mathrm{d}\xi}.\label{eq:dw}
\end{align}
Here the abbreviation $\mu(\xi,v)\equiv (\xi-v)/(1-\xi v)$ is introduced for the fluid velocity $v$ seen in a local frame comoving with the velocity $\xi$. The underlying microphysics is encoded in the sound velocity $c_s^2=\partial_\xi p/\partial_\xi e$. In general, sound velocities $c_+^2=\partial_\xi p_+/\partial_\xi e_+$ and $c_-^2=\partial_\xi p_-/\partial_\xi e_-$ in false and true vacua can be different constants from the bag model with $c_s^2=1/3$.

To further maintain the EOM $\nabla_\mu T^{\mu\nu}=0$ also across a discontinuous interface, one can impose the junction conditions across the bubble wall,
\begin{align}
\partial_\mu T^{\mu0}&=\partial_{\bar{t}}T^{\bar{t}\bar{t}}+\partial_{\bar{z}}T^{\bar{z}\bar{t}}=\partial_{\bar{z}}T^{\bar{z}\bar{t}}=0\nonumber\\
&\Rightarrow w_-\bar{\gamma}_-^2\bar{v}_-=w_+\bar{\gamma}_+^2\bar{v}_+,\label{eq:Wall1stJunction}\\
\partial_\mu T^{\mu1}&=\partial_{\bar{t}}T^{\bar{t}\bar{z}}+\partial_{\bar{z}}T^{\bar{z}\bar{z}}=\partial_{\bar{z}}T^{\bar{z}\bar{z}}=0\nonumber\\
&\Rightarrow w_-\bar{\gamma}_-^2\bar{v}_-^2+p_-=w_+\bar{\gamma}_+^2\bar{v}_+^2+p_+,\label{eq:Wall2ndJunction}
\end{align}
and the junction conditions across the shock-wave front (if any),
\begin{align}
\partial_\mu T^{\mu0}&=\partial_{\tilde{t}}T^{\tilde{t}\tilde{t}}+\partial_{\tilde{z}}T^{\tilde{z}\tilde{t}}=\partial_{\tilde{z}}T^{\tilde{z}\tilde{t}}=0\nonumber\\
&\Rightarrow w_L\tilde{\gamma}_L^2\tilde{v}_L=w_R\tilde{\gamma}_R^2\tilde{v}_R,\label{eq:Shock1stJunction}\\
\partial_\mu T^{\mu1}&=\partial_{\tilde{t}}T^{\tilde{t}\tilde{z}}+\partial_{\tilde{z}}T^{\tilde{z}\tilde{z}}=\partial_{\tilde{z}}T^{\tilde{z}\tilde{z}}=0\nonumber\\
&\Rightarrow w_L\tilde{\gamma}_L^2\tilde{v}_L^2+p_L=w_R\tilde{\gamma}_R^2\tilde{v}_R^2+p_R,\label{eq:Shock2ndJunction}
\end{align}
where $w_\pm$, $\bar{v}_\pm$, and $\bar{\gamma}_\pm\equiv\gamma(\bar{v}_\pm)$ are the enthalpy, wall-frame fluid velocity, and corresponding Lorentz factors just in front and back of the bubble wall, respectively, while $w_{L/R}$, $\tilde{v}_{L/R}$, and $\tilde{\gamma}_{L/R}\equiv\gamma(\tilde{v}_{L/R})$ are the enthalpy, shock-frame fluid velocity, and corresponding Lorentz factors just in back and in front of the shock-wave front, respectively. The fluid velocity and enthalpy profiles can be solved via hydrodynamic equations~\eqref{eq:dv} and~\eqref{eq:dw} provided with junction conditions~\eqref{eq:Wall1stJunction} and~\eqref{eq:Shock1stJunction} at the bubble wall and shock-wave front (if any), respectively.

\subsection{General backreaction force}\label{subsec:GeneralBackreaction}

After reducing the scalar-plasma system into a wall-fluid system, both the EOMs~\eqref{eq:EOMscalar} and~\eqref{eq:EOMplasma} can be further simplified by projecting them along the direction of expansion, leading to a balance equation for the scalar-wall expansion and a violation equation for the enthalpy-flow conservation, respectively. As we will show shortly below, combining these two equations of scalar-wall expansion and enthalpy-flow violation, we can derive a general formula for the backreaction force purely from the perfect-fluid hydrodynamics alone without reference to the underlying microscopic particle physics necessary in solving the Boltzmann equation. In the next section, we will apply our hydrodynamic backreaction force to a general EOS beyond the simple bag model and then prove its balance to the driving force analytically in the section after that.

We first turn to project the scalar EOM~\eqref{eq:EOMscalar} along the direction of bubble expansion by integrating it along with the wall derivative over the self-similar coordinate as
\begin{equation}
\int_0^1\mathrm{d}\xi\frac{\mathrm{d}\phi}{\mathrm{d}\xi}\left(\nabla^2\phi-\frac{\partial V_\mathrm{eff}}{\partial\phi}+\frac{\partial p_{\delta f}}{\partial\phi}\right)=0.
\end{equation}
The first term as a total derivative of $\phi'(\xi)^2/2$  can be fully integrated out as $\phi'(\xi=1)/2-\phi'(\xi=0)/2=0$ in the thin-wall limit. The second term belongs to a total derivative term subtracted by a temperature jumping part as $(\partial V_\mathrm{eff}/\partial\phi)(\mathrm{d}\phi/\mathrm{d}\xi)=\mathrm{d}V_\mathrm{eff}/\mathrm{d}\xi-(\partial V_\mathrm{eff}/\partial T)(\mathrm{d}T/\mathrm{d}\xi)$. Hence, we arrive at a balance equation~\cite{Wang:2022txy},
\begin{align}
\label{eq:balance}
p_\mathrm{dr}\equiv \Delta V_\mathrm{eff}&=\int_0^1\mathrm{d}\xi \frac{\mathrm{d}T}{\mathrm{d}\xi}\frac{\partial V_\mathrm{eff}}{\partial T}+\int_0^1\mathrm{d} \xi \frac{\mathrm{d}\phi}{\mathrm{d}\xi}\frac{\partial p_{\delta f}}{\partial \phi}\nonumber\\
&\equiv p_\mathrm{th}+p_\mathrm{fr}\equiv p_\mathrm{br},
\end{align}
between the driving force (per unit area) $p_\mathrm{dr}\equiv\Delta V_\mathrm{eff}$ and the backreaction force (per unit area) $p_\mathrm{br}= p_\mathrm{th}+p_\mathrm{fr}$ consisting of a thermal gradient force (per unit area) $p_\mathrm{th}=\int\mathrm{d}T\,\partial_T V_\mathrm{eff}$ and a nonequilibrium friction force (per unit area) $p_\mathrm{fr}=\int\mathrm{d}\phi\,\partial_\phi p_{\delta f}$. Note that the friction force defined here is different from what is usually called ``friction force'' in the literature by computing the particle momentum recoils on the bubble wall, which is closer to our total backreaction force $p_\mathrm{br}$ acting on the wall alone, $p_\mathrm{br}|_\mathrm{wall}$. Besides the wall contribution, the full backreaction force also receives contributions from the sound shell, $p_\mathrm{br}|_\mathrm{shell}$, and shock-wave front (if any), $p_\mathrm{br}|_\mathrm{shock}$. As we will see in the next section, the backreaction force $p_\mathrm{br}|_\mathrm{wall}$ acting on the wall alone is not sufficient to balance the driving force $p_\mathrm{dr}$, and we need the full knowledge of the backreaction force $p_\mathrm{br}=p_\mathrm{br}|_\mathrm{wall}+p_\mathrm{br}|_\mathrm{shell}+p_\mathrm{br}|_\mathrm{shock}$. Remarkably, this can be done purely from the perfect-fluid hydrodynamics alone without going to the details of underlying particle physics.

We next turn to project the plasma EOM~\eqref{eq:EOMplasma} along the direction of bubble expansion by contracting it with the plasma fluid velocity as
\begin{equation}\label{eq:projection}
u_{\nu}\nabla_{\mu}(wu^{\mu}u^{\nu}+p_f\eta^{\mu\nu})+u_{\nu}\nabla^{\nu}\phi \frac{\partial \Delta V_T}{\partial \phi}=u_{\nu}\nabla^{\nu}\phi \frac{\partial p_{\delta f}}{\partial \phi},
\end{equation}
where we have used $w=w_\phi+w_f=w_f=e_f+p_f$ in the perfect-fluid energy-momentum tensor~\eqref{eq:plasmafluid}. After using the contraction relations $u_{\nu}u^{\nu}=-1$ and  $u_{\nu}\nabla_{\mu}u^{\nu}=0$ as well as the total derivative $\nabla_{\mu}p_f=-\nabla_{\mu}\Delta V_T=-\nabla_{\mu}T(\partial\Delta V_{T}/\partial T)-\nabla_{\mu}\phi(\partial\Delta V_{T}/\partial \phi)=-\nabla_{\mu}T(\partial V_\mathrm{eff}/\partial T)-\nabla_{\mu}\phi(\partial\Delta V_{T}/\partial \phi)$, we finally arrive at the violation equation for the enthalpy-flow conservation as
\begin{equation}
\label{eq:EnthalpyFlow}
-\nabla_{\mu}(wu^{\mu})=u^{\mu}\nabla_{\mu}T\frac{\partial V_\mathrm{eff}}{\partial T}+u^{\mu}\nabla_{\mu}\phi \frac{\partial p_{\delta f}}{\partial \phi}.
\end{equation}
Note that for scalar functions $F(\xi)=T(\xi), \phi(\xi)$, we can write $u^{\mu}\nabla_{\mu}F=(\gamma,\gamma v,0,0) \cdot (-\xi/t,1/t,0,0)\partial_{\xi}F=(\gamma/t)(v-\xi)F'(\xi)$ explicitly in the self-similar coordinate,
\begin{equation}
    \label{enthalpy-flow-new}
    \frac{t\nabla_{\mu}(wu^{\mu})}{\gamma (\xi-v)}=\frac{\text{d}T}{\text{d}\xi}\frac{\partial V_{\rm eff}}{\partial T}+\frac{\text{d}\phi}{\text{d}\xi} \frac{\partial p_{\delta f}}{\partial \phi}.
\end{equation}
Note also that the right-hand side of above expression exactly reproduces the integrand of the backreaction force~\eqref{eq:balance}, and thus, we can obtain a preliminary evaluation of the total backreaction force as
\begin{equation} \label{eq:p_br}
p_\mathrm{br}=\int_0^1\mathrm{d}\xi \frac{t\nabla_{\mu}(wu^{\mu})}{\gamma (\xi-v)}.
\end{equation}
This expression is not ready for a hydrodynamic evaluation yet. To actually evaluate the total backreaction force, we can explicitly expand $\nabla_{\mu}(wu^{\mu})$ in the plasma frame with the self-similar coordinate as
\begin{equation}
\nabla_{\mu}(wu^{\mu})=\frac{\gamma}{t}(v-\xi)\partial_{\xi}w+\frac{D}{\xi}\frac{\gamma}{t}wv+\frac{\gamma^3}{t}w(1-\xi v)\partial_{\xi}v,
\end{equation}
where $D=0,1,2$ are for the bubble expansion of planar, cylindrical, and spherical wall geometries, respectively. Now we finally derive our hydrodynamic evaluation of the total backreaction force (per unit area)~\cite{Wang:2022txy} as
\begin{equation}
\label{eq:brhydro}
p_\mathrm{br}=\int_{0}^{1} \text{d}\xi \left( -\frac{\text{d}w}{\text{d}\xi}+\frac{Dwv}{\xi(\xi-v)}+\frac{w\gamma^2}{\mu(\xi,v)}\frac{\text{d}v}{\text{d}\xi}\right).
\end{equation}

The hydrodynamic evaluation of the total backreaction force~\eqref{eq:brhydro} can be further split into the sound-shell part, bubble-wall part, and shock-wave-front part (if any)~\cite{Wang:2022txy},
\begin{align}
p_\mathrm{br}|_\mathrm{shell}&=-\int_\mathrm{shell}\mathrm{d}\xi\frac{\mathrm{d}w}{\mathrm{d}\xi}\frac{c_s^2}{1+c_s^2},\label{eq:Fbackhydroshell}\\
p_\mathrm{br}|_\mathrm{wall}&=-\overline{\Delta} w+\int_{v(\xi_w^-)}^{v(\xi_w^+)}\mathrm{d}v\frac{w(v)\gamma(v)^2}{\mu(\xi_w,v)},\label{eq:Fbackhydrowall}\\
p_\mathrm{br}|_\mathrm{shock}&=-\widetilde{\Delta} w+\int_{v(\xi_{sh}^-)}^{v(\xi_{sh}^+)}\mathrm{d}v\frac{w(v)\gamma(v)^2}{\mu(\xi_{sh},v)}\label{eq:Fbackhydroshock},
\end{align}
where $v(\xi_w^\pm)\equiv v_\pm$ are fluid velocities in the background plasma frame just in the front and the back of the bubble-wall $\xi_w$, respectively, while $v(\xi_{sh}^\pm)$ are fluid velocities in the background plasma frame just in the front and the back of the shock-wave-front $\xi_{sh}$, respectively. Here the sound-shell contribution is obtained from~\eqref{eq:brhydro} by replacing $v/\xi$ and $\mathrm{d}v/\mathrm{d}\xi$ terms with hydrodynamic EOMs~\eqref{eq:dv} and~\eqref{eq:dw}, and the integral is only implemented over the continuous parts of the sound-shell without crossing any discontinuous interfaces. On the other hand, the discontinuous contributions from the bubble wall and shock-wave front (if any) contain not only the enthalpy difference across the bubble wall and shock-wave front,
\begin{align}
\overline{\Delta} w&\equiv\lim_{\delta\to0^+}w(\xi_w+\delta)-\lim_{\delta\to0^-}w(\xi_w+\delta)\nonumber\\
&\equiv w(\xi_w^+)-w(\xi_w^-)\equiv w_+-w_-,\label{eq:diffbar}\\
\widetilde{\Delta} w&\equiv\lim_{\delta\to0^+}w(\xi_{sh}+\delta)-\lim_{\delta\to0^-}w(\xi_{sh}+\delta)\nonumber\\
&\equiv w(\xi_{sh}^+)-w(\xi_{sh}^-)\equiv w_R-w_L,\label{eq:difftilde}
\end{align}
but also a nontrivial integral from the last term of~\eqref{eq:brhydro}, which should be calculated using a mathematical trick introduced in Ref.~\cite{Wang:2022txy}. The difficulty of integrating the last term in~\eqref{eq:brhydro} is that both $w$ and $v$ are discontinuous functions across the bubble wall/shock-wave front (if any). To make a mathematically consistent integration, one would require extra input from some continuous function $w(v)$ that could relate $w$ and $v$ across the discontinuous interfaces. Obviously, this can be achieved from the junction conditions~\eqref{eq:Wall1stJunction} and~\eqref{eq:Shock1stJunction}. The validity of this mathematical trick has been confirmed analytically when proving the exact balance between the driving force and our hydrodynamic evaluation on the backreaction force for the case with a bag EOS~\cite{Li:2023xto}. In the next section, we will apply our hydrodynamic backreaction force~\eqref{eq:brhydro} to a general EOS beyond the bag model, and then prove its balance to the driving force in the section after that.

\subsection{Alternative proof of general backreaction force}\label{subsec:ProofGeneralBackreaction}

It is worth noting that our expression~\eqref{eq:p_br} for the total backreaction force does not rely on whether to include a nonequilibrium pressure $\delta p$ to the definitions of the total pressure $p$ by hand as $p=-V_\mathrm{eff}+\delta p$ as usually done in Refs.~\cite{Ai:2021kak,Ai:2024shx}.

To see this,  note that Eq.~\eqref{eq:p_br} can also be derived directly from the conservation of the total energy-momentum tensor $T^{\mu\nu}$ that is assumed to be described by a perfect-fluid $T^{\mu\nu}=(e+p)u^\mu u^\nu+p\eta^{\mu\nu}$, whose conservation equation $\nabla_\mu T^{\mu\nu}=0$ after projected along $u_\nu$ leads to
\begin{align}
0 = u_\nu \nabla_\mu T^{\mu\nu} &= u_{\nu}\nabla_{\mu}(w u^{\mu}u^{\nu}+p \eta^{\mu\nu})\nonumber\\
& =  -\nabla_{\mu}(w u^{\mu}) + u^\mu\nabla_\mu p.
\end{align}
Since for any scalar functions $F(\xi)$, we
can expand $u^{\mu}\nabla_{\mu}F=(\gamma,\gamma v,0,0) \cdot (-\xi/t,1/t,0,0)\partial_{\xi}F=(\gamma/t)(v-\xi)F'(\xi)$ explicitly in the self-similar coordinate, and then the equation above can be rewritten as
\begin{align}
    -\frac{\mathrm{d} p}{\mathrm{d}\xi} = \frac{t \nabla_{\mu}(w u^{\mu})}{\gamma (\xi - v)}.
\end{align}
Integrating over $\xi$ between the bubble center $\xi=0$ and null infinity $\xi=1$, the left-hand side simply gives rise to $-p|_{\xi=0}^{\xi=1}= \Delta V_{\mathrm{eff}}\equiv V_\mathrm{eff}(\phi_+, T(\xi=1))-V_\mathrm{eff}(\phi_-,T(\xi=0))$ no matter which definitions ($p=-V_\mathrm{eff}$ or $p=-V_\mathrm{eff}+\delta p$) are used since the out-of-equilibrium contribution $\delta p$ to the total pressure should necessarily vanish far away from the bubble wall. Therefore, we always arrive at our expression~\eqref{eq:p_br} ,
\begin{align}
p_{\mathrm{br}}=p_{\mathrm{dr}} \equiv \Delta V_{\mathrm{eff}} = \int_0^1\mathrm{d}\xi \frac{t\nabla_{\mu}(wu^{\mu})}{\gamma (\xi-v)},
\end{align}
when the total backreaction force eventually balances the driving force $p_{\mathrm{br}}=p_{\mathrm{dr}}$ for a nonrunaway steady-state self-similar thin-wall expansion.

In fact, we need not bother to manually include nonequilibrium pressures $\delta p$ and $\delta p_f$ to the total pressure $p$ and fluid pressure $p_f$ by hand as $p=-V_\mathrm{eff}+\delta p$ and $p_f=-\Delta V_T+\delta p_f$, respectively, since the effective potential $V_\mathrm{eff}$ already contains all the necessary coupling terms that contribute to the collision terms on the right-hand side of the Boltzmann equation for the total distribution function. The nonequilibrium effect automatically manifests itself on the right-hand side of coupled EOMs~\eqref{eq:EOMscalar} and~\eqref{eq:EOMplasma}, which can be rewritten as
\begin{align}
\nabla_\mu\nabla^\mu\phi&=-\frac{\partial P}{\partial\phi},\\
\nabla_\mu T_f^{\mu\nu}&=\nabla^\nu\phi\frac{\partial P_f}{\partial\phi},
\end{align}
if one really wants to define some kind of a total pressure $P=-V_\mathrm{eff}+p_{\delta f}$ and its fluid contribution $P_f=-\Delta V_ T+p_{\delta f}$ when including the nonequilibrium effect. Hence, we naturally reveal the nonequilibrium contribution to the pressure $\delta p_f$ as the nonequilibrium contribution $p_{\delta f}$ to the transfer flow $f^\mu$ in Eq.~\eqref{eq:TransferFlowAnsatz}. Note that this newly defined $P$ and $P_f$ are not the ones that appear in the energy-momentum tensors.

Even if we start from the traditional definitions $p=-V_\mathrm{eff}+p_{\delta f}$ and $p_f=-\Delta V_T+p_{\delta f}$ with the nonequilibrium pressure $p_{\delta f}$ induced by the nonequilibrium part of the distribution function, we can still project the plasma EOM~\eqref{eq:EOMplasma} along the direction of the bubble expansion by contracting it with the plasma fluid velocity as done for Eq.~\eqref{eq:projection},  whose left-hand side reads
\begin{align}
&u_\nu\nabla_\mu(wu^\mu u^\nu+p_f\eta^{\mu\nu})+u_\nu\nabla^\nu\phi\frac{\partial\Delta V_T}{\partial\phi}\nonumber\\
&=-\nabla_\mu(wu^\mu)+u^\mu\nabla_\mu p_f+u_\nu\nabla^\nu\phi\frac{\partial\Delta V_T}{\partial\phi}\nonumber\\
&=-\nabla_\mu(wu^\mu)+u_\nu\nabla^\nu(-\Delta V_T+p_{\delta f})+u_\nu\nabla^\nu\phi\frac{\partial\Delta V_T}{\partial\phi}\nonumber\\
&=-\nabla_\mu(wu^\mu)-u_\nu\left(\nabla^\nu T\frac{\partial\Delta V_T}{\partial T}+\nabla^\nu\phi\frac{\partial\Delta V_T}{\partial\phi}\right)\nonumber\\
&\quad+u_\nu\nabla^\nu p_{\delta f}+u_\nu\nabla^\nu\phi\frac{\partial\Delta V_T}{\partial\phi}\nonumber\\
&=-\nabla_\mu(wu^\mu)-u_\nu\nabla^\nu T\frac{\partial\Delta V_T}{\partial T}+u_\nu\nabla^\nu p_{\delta f}.
\end{align}
Substituting this equation into Eq.~\eqref{eq:projection} and then noting that $u_\nu\nabla^\nu p_{\delta f}=u_\nu\left(\nabla^\nu T\frac{\partial p_{\delta f}}{\partial T}+\nabla^\nu\phi\frac{\partial p_{\delta f}}{\partial\phi}\right)$, we have
\begin{align}
-\nabla_\mu(wu^\mu)=u_\nu\nabla^\nu T\frac{\partial\Delta V_T}{\partial T}-u_\nu\nabla^\nu T\frac{\partial p_{\delta f}}{\partial T},
\end{align}
which is exactly the enthalpy-flow equation~\eqref{eq:EnthalpyFlow} up to a total derivative (also note that $\frac{\partial\Delta V_T}{\partial T}=\frac{\partial V_\mathrm{eff}}{\partial T}$),
\begin{align}
-\nabla_\mu(wu^\mu)=u_\nu\nabla^\nu T\frac{\partial V_\mathrm{eff}}{\partial T}+u_\nu\nabla^\nu\phi\frac{\partial p_{\delta f}}{\partial\phi}-u_\nu\nabla^\nu p_{\delta f}.
\end{align}
After being expressed in the self-similar coordinate
\begin{align}
\frac{t\nabla_\mu(wu^\mu)}{\gamma(\xi-v)}=\frac{\mathrm{d}T}{\mathrm{d}\xi}\frac{\partial V_\mathrm{eff}}{\partial T}+\frac{\mathrm{d}\phi}{\mathrm{d}\xi}\frac{\partial p_{\delta f}}{\partial\phi}-\frac{\mathrm{d}p_{\delta f}}{\mathrm{d}\xi},
\end{align}
the right-hand side also reproduces the integrand of the backreaction force~\eqref{eq:balance} up to a total derivative, and hence the same expression for the total backreaction force~\eqref{eq:p_br} can be derived,
\begin{align}
p_\mathrm{br}=\int_0^1\mathrm{d}\xi\frac{t\nabla_\mu(wu^\mu)}{\gamma(\xi-v)}-p_{\delta f}\bigg|_0^1,
\end{align}
since the nonequilibrium contribution to the pressure should vanish in the innermost and outermost ends of the bubble profile, $p_{\delta f}(\xi=0,1)=0$.

\section{Backreaction force beyond bag EOS}\label{sec:pbrmuModel}

The fluid velocity profile $v(\xi)$ can be solved from the hydrodynamic EOM~\eqref{eq:dv} for a particular expansion mode with corresponding junction condition(s)~\eqref{eq:Wall1stJunction} and/or~\eqref{eq:Shock1stJunction} given a specific EOS. In our previous studies~\cite{Wang:2022txy,Li:2023xto}, the MIT bag EOS~\cite{Chodos:1974je} is adopted by parametrizing the energy density $e_{\pm}=a_{\pm} T^{4}_{\pm} + V_0^{\pm}$ and pressure $p_{\pm}=\frac{1}{3}a_{\pm}T^4_{\pm}-V_0^{\pm}$ as simple collections of vacuum energy $V_0^{\pm}\equiv V_0(\phi_{\pm})$ and ideal gas with the effective number of relativistic degrees of freedom $a_{\pm}\equiv (\pi^2/30)g_{\rm eff}^{\pm}$ in the symmetric and broken phases, respectively. Thus, the sound velocity $c_s^2=\partial p/\partial e=1/3$ is a single constant throughout the bubble for the bag EOS. In general, the sound velocity $c_s(\xi)$ should also vary with $\xi$~\cite{Wang:2022lyd}, and if we neglect the changes inside the sound-shell, it can be approximated with two different constants $c_-$ and $c_+$ inside and outside of the bubble, respectively, which can be achieved in the so-called $\nu$-model with the energy density and pressure parametrized via $\nu_{\pm}\equiv 1+1/c_{\pm}^2$ as~\cite{Leitao:2014pda}
\begin{align}
e_{\pm}&=a_{\pm}T_{\pm}^{\nu_{\pm}}+V_0^{\pm},\label{eq:nue}\\
p_{\pm}&=c_{\pm}^2a_{\pm}T_{\pm}^{\nu_{\pm}}-V_0^{\pm}.\label{eq:nup}
\end{align}
Inserting the above EOS into the junction conditions~\eqref{eq:Wall1stJunction} and~\eqref{eq:Wall2ndJunction} gives rise to a relation between the wall-frame fluid velocities $\bar{v}_\pm$ near the wall as~\cite{Leitao:2014pda}
\begin{equation}
\overline{v}_{+}=\frac{1}{1+\alpha_{+}}\left( qX_{+}\pm \sqrt{q^2X_{-}^2+\alpha_{+}^2+(1-c_{+}^2)\alpha_+ +q^2c_{-}^2-c_{+}^2} \right),
\label{eq:vbarpm}
\end{equation}
where $q\equiv(1+c_{+}^2)/(1+c_{-}^2)$, $X_{\pm}\equiv \overline{v}_{-}/2\pm c_{-}^2/(2\overline{v}_{-})$, and $\alpha_{+}\equiv \Delta V_0 /(a_+ T_+^{\nu_+})$ is the strength factor just in front of the bubble wall, while $\alpha_{N}\equiv \Delta V_0 /(a_+ T_N^{\nu_+})$ is the strength factor in null infinity at $\xi=1$. It is easy to see that $\alpha_{+}w_{+}=\alpha_{N}w_{N}=(1+c_{+}^2)\Delta V_0$ with $\Delta V_0 \equiv V_0^{+}-V_0^{-}$. Equipped with the above $\nu$-model EOS, the hydrodynamic EOMs~\eqref{eq:dv} and~\eqref{eq:dw} can be solved with the junction condition(s)~\eqref{eq:Wall1stJunction} and/or~\eqref{eq:Shock1stJunction} for four different expansion modes: weak detonation, weak deflagration, Jouguet deflagration, and Jouguet detonation. Now we apply our hydrodynamic evaluation of backreaction force~\eqref{eq:brhydro} to this $\nu$-model EOS along with explicit decomposition from~\eqref{eq:Fbackhydroshell}, \eqref{eq:Fbackhydrowall}, and~\eqref{eq:Fbackhydroshock}.

\subsection{Weak detonation}\label{subsec:WeakDetona}

For the weak detonation mode, the continuous part of the fluid velocity profile is the sound-shell regime between $c_-<\xi<\xi_w$ where the sound velocity $c_s$ in~\eqref{eq:Fbackhydroshell} takes the value $c_-$; hence, the sound-shell part of the full backreaction force reads
\begin{equation}
\label{eq:DetonaShell}
p_\mathrm{br}|_\mathrm{shell}=-\int_{c_-}^{\xi_w^-}\mathrm{d}\xi\frac{\mathrm{d}w}{\mathrm{d}\xi} \frac{c_-^2}{1+c_-^2}=-\frac{c_-^2}{1+c_-^2}(w_--w_s),
\end{equation}
where $w_-\equiv w(\xi_w^-\equiv\xi_w+0^-)$ is the enthalpy just behind the wall, while the enthalpy $w_s\equiv w(\xi=c_-+0^\pm)$ is continuous at $\xi=c_-$ and equals the enthalpy $w_s\equiv w(\xi=c_-)=w(\xi=0)\equiv w_O$ at the origin (namely bubble center). On the other hand, the discontinuous part of the fluid velocity profile is at the bubble wall $\xi=\xi_w$ for the weak detonation, and hence, the discontinuous part of the full backreaction force can be calculated from the bubble-wall contribution~\eqref{eq:Fbackhydrowall} as
\begin{align}
\label{eq:DetonaWall}
p_\mathrm{br}|_\mathrm{wall}&=-(w_+-w_-)+\int_{v_-}^{v_+}\mathrm{d}v\frac{w(v)\gamma(v)^2}{\mu(\xi_w,v)}\nonumber\\
&=-(w_+-w_-)+\frac{v_-}{v_--\xi_w}w_+,
\end{align}
where $v_-\equiv v(\xi=\xi_w+0^-)$ is the plasma-frame fluid velocity just behind the wall, while the enthalpy just in front of the wall $w_+\equiv w(\xi=\xi_w+0^+)=w(\xi=1)\equiv w_N$ equals the enthalpy at null infinity. In calculating the integral in the bubble-wall contribution, we have used the fact that the fluid velocity in front of the wall $v_+=0$ is static for the weak detonation, and the abbreviation $\gamma(v)$ reads $(1-v^2)^{-1/2}$ hereafter, while $w(v)$ is a continuous function relating $v_\pm$ and $w_\pm$ across the bubble wall from the junction condition~\eqref{eq:Wall1stJunction},
\begin{equation}
w(v)=w_+\frac{\bar{\gamma}_{+}^2\bar{v}_+}{\bar{\gamma}^2\bar{v}}=w_+\frac{\xi_w}{1-\xi_w^2}\frac{1-\mu(\xi_w,v)^2}{\mu(\xi_w,v)},
\end{equation}
where we have used $\bar{v}_+=\xi_w$ and $\bar{v}=\mu(\xi_w,v)$. Therefore, the full backreaction force reads 
\begin{equation}
\label{eq:DetonaFull}
p_\mathrm{br}=-(w_N-w_-)+\frac{v_-}{v_--\xi_w}w_N-\frac{c_-^2}{1+c_-^2}(w_--w_O).
\end{equation}

\subsection{Weak deflagration}\label{subsec:WeakDeflag}

For the weak deflagration, the continuous part of the fluid velocity profile is the sound-shell regime between $\xi_w<\xi<\xi_{sh}$ where the sound velocity $c_s$ in~\eqref{eq:Fbackhydroshell} takes the values $c_+$; hence, the sound-shell part of the full backreaction force reads
\begin{equation}
\label{eq:DeflagShell}
p_\mathrm{br}|_\mathrm{shell}=-\int_{\xi_w^+}^{\xi_{sh}^-}\mathrm{d}\xi\frac{\mathrm{d}w}{\mathrm{d}\xi} \frac{c_+^2}{1+c_+^2}=-\frac{c_+^2}{1+c_+^2}(w_L-w_+),
\end{equation}
where $w_L\equiv w(\xi_{sh}^-\equiv\xi_{sh}+0^-)$ is the enthalpy just behind the shock-wave front, while $w_+\equiv w(\xi_w^+\equiv\xi_w+0^+)$ is the enthalpy just in front of the wall. On the other hand, the discontinuous parts of the fluid velocity profile are at the bubble-wall $\xi_w$ and shock-wave-front $\xi_{sh}$ for the weak deflagration, and hence, the corresponding bubble-wall and shock-wave-front parts of the full backreaction force can be calculated from~\eqref{eq:Fbackhydrowall} and~\eqref{eq:Fbackhydroshock} as
\begin{align}
p_\mathrm{br}|_\mathrm{wall}&=-(w_+-w_-)+\int_{0}^{v_+}\mathrm{d}v\frac{w(v)\gamma(v)^2}{\mu(\xi_w,v)}\nonumber\\
&=-(w_+-w_-)-\frac{v_+}{v_+-\xi_w}w_-,\label{eq:DeflagWall}\\
p_\mathrm{br}|_\mathrm{shock}&=-(w_R-w_L)+\int_{v_{sh}}^{0}\mathrm{d}v\frac{w(v)\gamma(v)^2}{\mu(\xi_{sh},v)}\nonumber\\
&=-(w_R-w_L)+\frac{v_{sh}}{v_{sh}-\xi_{sh}}w_R,\label{eq:DeflagShock}
\end{align}
respectively, where $v_+=v(\xi_w+0^+)$ and $v_{sh}=v(\xi_{sh}+0^-)$ are plasma-frame fluid velocities just in front of the bubble wall and just behind the shock-wave front, respectively, and the enthalpy just behind the wall $w_-\equiv w(\xi_w+0^-)=w(\xi=0)\equiv w_O$ equals the enthalpy at the origin, while the enthalpy just in front of the shock-wave-front $w_R\equiv w(\xi_{sh}+0^+)=w(\xi=1)\equiv w_N$ equals to the enthalpy at null infinity. In calculating the integrals in the bubble-wall and shock-wave-front contributions, we have used the facts that both fluid velocities behind the bubble-wall $v_-=0$ and in front of the shock-wave-front $v(\xi_{sh}+0^+)=0$ are static for the weak deflagration. Here $w(v)$ in the integral of the bubble-wall contribution is a continuous function relating $v_\pm$ and $w_\pm$ across the bubble wall from the junction condition~\eqref{eq:Wall1stJunction}, while $w(v)$ in the integral of the shock-wave-front contribution is also a continuous function relating  $v(\xi_{sh}+0^\pm)$ and $w_{R/L}$ across the shock-wave front from the junction condition~\eqref{eq:Shock1stJunction}, that is,
\begin{align}
w(v)|_\mathrm{wall}&=w_-\frac{\bar{\gamma}_{-}^2\bar{v}_-}{\bar{\gamma}^2\bar{v}}=w_-\frac{\xi_w}{1-\xi_w^2}\frac{1-\mu(\xi_w,v)^2}{\mu(\xi_w,v)},\\
w(v)|_\mathrm{shock}&=w_R\frac{\tilde{\gamma}_{R}^2\tilde{v}_R}{\tilde{\gamma}^2\tilde{v}}=w_R\frac{\xi_{sh}}{1-\xi_{sh}^2}\frac{1-\mu(\xi_{sh},v)^2}{\mu(\xi_{sh},v)},
\end{align}
where we have used $\bar{v}_-=\xi_w$, $\bar{v}=\mu(\xi_w,v)$, $\tilde{v}_R=\xi_{sh}$, and $\tilde{v}=\mu(\xi_{sh},v)$. Therefore, the full backreaction force reads
\begin{align}
\label{eq:DeflagFull}
p_\mathrm{br}&=-\frac{c_+^2}{1+c_+^2}(w_L-w_+)-(w_+-w_O)\nonumber\\
&-\frac{v_+}{v_+-\xi_w}w_O-(w_N-w_L)+\frac{v_{sh}}{v_{sh}-\xi_{sh}}w_N.
\end{align}

\subsection{Jouguet detonation and deflagration}\label{subsec:Jouguet}

For the Jouguet detonation and Jouguet deflagration, the fluid velocity profile admits two continuous parts separated by the bubble wall $\xi=\xi_w$: the inner sound-shell between $c_-<\xi<\xi_w$ and the outer sound-shell between $\xi_w<\xi<\xi_{sh}$, where the corresponding sound velocities in~\eqref{eq:Fbackhydroshell} take the values of $c_-$ and $c_+$, respectively. Hence, the sound-shell contribution to the full backreaction force reads
\begin{align}
\label{eq:JouguetShell}
p_\mathrm{br}|_\mathrm{shell}&=-\left(\int_{c_-}^{\xi_w^-}+\int_{\xi_w^+}^{\xi_{sh}^-}\right)\mathrm{d}\xi\frac{\mathrm{d}w}{\mathrm{d}\xi} \frac{c_s^2}{1+c_s^2}\nonumber\\
&=-\frac{c_-^2}{1+c_-^2}(w_--w_s)-\frac{c_+^2}{1+c_+^2}(w_L-w_+),
\end{align}
where $w_s\equiv w(c_-+0^\pm)=w(\xi=0)\equiv w_O$, $w_-\equiv w(\xi_w^-\equiv\xi_w+0^-)$, $w_+\equiv w(\xi_w^+\equiv\xi_w+0^+)$, and $w_L\equiv w(\xi_{sh}^-\equiv\xi_{sh}+0^-)$. On the other hand, the discontinuous parts of the fluid velocity profile are at the bubble-wall $\xi_w$ and shcok-wave-front $\xi_{sh}$ for both Jouguet detonation and Jouguet deflagration, and hence, the corresponding bubble-wall and shock-wave-front contributions to the full backreaction force can be calculated from~\eqref{eq:Fbackhydrowall} and~\eqref{eq:Fbackhydroshock} as
\begin{align}
p_\mathrm{br}|_\mathrm{wall}&=-(w_+-w_-)+\int_{v_-}^{v_+}\text{d}v\frac{w(v)\gamma(v)^2}{\mu(\xi_w,v)}\nonumber\\
&=-(w_+-w_-)+\frac{c_-(1-\xi_w^2)(v_--v_+)w_-}{(1-c_-^2)(v_--\xi_w)(\xi_w-v_+)},\label{eq:JouguetWall}\\
p_\mathrm{br}|_\mathrm{shock}&=-(w_R-w_L)+\int_{v_{sh}}^{0}\text{d}v\frac{w(v)\gamma(v)^2}{\mu(\xi_{sh},v)}\nonumber\\
&=-(w_R-w_L)+\frac{v_{sh}}{v_{sh}-\xi_{sh}}w_R,\label{eq:JouguetShock}
\end{align}
respectively, where $v_+\equiv v(\xi_w+0^+)$, $v_-\equiv v(\xi_w+0^-)$, $v_{sh}\equiv v(\xi_{sh}+0^-)$, and $w_R\equiv w(\xi_{sh}^+\equiv\xi_{sh}+0^+)=w(\xi=1)\equiv w_N$. In calculating the integrals in the bubble-wall and shock-wave-front contributions, we have used the fact that $v(\xi_{sh}+0^+)=0$, and $w(v)$ in the integral of the bubble-wall contribution is a continuous function relating $v_\pm$ and $w_\pm$ across the bubble wall from the junction condition~\eqref{eq:Wall1stJunction}, while $w(v)$ in the integral of the shock-wave-front contribution is also a continuous function relating  $v(\xi_{sh}+0^\pm)$ and $w_{R/L}$ across the shock-wave front from the junction condition~\eqref{eq:Shock1stJunction}, that is,
\begin{align}
w(v)|_\mathrm{wall}&=w_-\frac{\bar{\gamma}_{-}^2\bar{v}_-}{\bar{\gamma}^2\bar{v}}=w_-\frac{c_-}{1-c_-^2}\frac{1-\mu(\xi_w,v)^2}{\mu(\xi_w,v)},\\
w(v)|_\mathrm{shock}&=w_R\frac{\tilde{\gamma}_{R}^2\tilde{v}_R}{\tilde{\gamma}^2\tilde{v}}=w_R\frac{\xi_{sh}}{1-\xi_{sh}^2}\frac{1-\mu(\xi_{sh},v)^2}{\mu(\xi_{sh},v)},
\end{align}
where we have used $\bar{v}_-=c_-$, $\bar{v}=\mu(\xi_w,v)$, $\tilde{v}_R=\xi_{sh}$, and $\tilde{v}=\mu(\xi_{sh},v)$. Therefore, the full backreaction force consisting of inner shell, bubble wall, outer shell, and shock front reads
\begin{align}
\label{eq:JouguetFull}
p_\mathrm{br}=&-\frac{c_-^2}{1+c_-^2}(w_--w_O)-(w_+-w_-)\nonumber\\
&+\frac{c_-(1-\xi_w^2)(v_--v_+)w_-}{(1-c_-^2)(v_--\xi_w)(\xi_w-v_+)}-\frac{c_+^2}{1+c_+^2}(w_L-w_+)\nonumber\\
&-(w_N-w_L)+\frac{v_{sh}}{v_{sh}-\xi_{sh}}w_N.
\end{align}

\section{Analytic consistency checks beyond bag EOS}\label{sec:ProofBalance}

To analytically verify our hydrodynamic evaluation on the backreaction force, we provide in this section two consistency checks. The first one is to check if the bubble-wall contribution to the hydrodynamic backreaction force could reproduce the pressure difference near the bubble wall from the junction condition~\eqref{eq:Wall2ndJunction},
\begin{equation}
\label{eq:brwalljunction}
p_\mathrm{br}|_\mathrm{wall}=w_{+}\bar{\gamma}_{+}^2\bar{v}_{+}^2-w_{-}\bar{\gamma}_{-}^2\bar{v}_{-}^2\equiv \overline{\Delta}(w\bar{\gamma}^2\bar{v}^2).
\end{equation}
A second check is to see if the full hydrodynamic backreaction force could exactly balance the driving force $p_\mathrm{dr}=\Delta V_\mathrm{eff}=-\Delta p=p_O-p_N$ for the $\nu$-model EOS~\eqref{eq:nue} and~\eqref{eq:nup},
\begin{equation}
\label{eq:drnumodel}
p_\mathrm{dr}=\frac{c_{-}^2}{1+c_{-}^2}w_{O}-\frac{c_{+}^2}{1+c_{+}^2}w_{N}+\frac{1}{1+c_{+}^2}\alpha_{+}w_{+},
\end{equation}
where $p_O\equiv p(\xi=0)=c_-^2a_-T_O^{\nu_-}-V_0^-$ and $p_N\equiv p(\xi=1)=c_+^2a_+T_N^{\nu_+}-V_0^+$ are the innermost and outermost pressures, respectively, while $w_O\equiv w(\xi=0)=(1+c_-^2)a_-T_O^{\nu_-}$ and $w_N\equiv w(\xi=1)=(1+c_+^2)a_+T_N^{\nu_+}$ are the innermost and outermost enthalpies, respectively. $\alpha_+\equiv (1+c_+^2)\Delta V_0/w_+$ is the strength factor just in front of the bubble wall and can be related to the strength factor $\alpha_N\equiv(1+c_+^2)\Delta V_0/w_N$ at the null infinity via $\alpha_+w_+=\alpha_N w_N=(1+c_+^2)\Delta V_0$. As we will see shortly below, proving this force balance is equivalent to pick the physical branch of the hydrodynamic solutions~\eqref{eq:vbarpm}, which is the plus-sign branch for both weak and Jouguet detonation modes, and the minus-sign branch for both weak and Jouguet deflagration modes.

\subsection{Weak detonation}\label{subsec:ProveWeakDetona}

For the weak detonation mode, recall that our hydrodynamic evaluation of the full backreaction force $p_\mathrm{br}=p_\mathrm{br}|_\mathrm{shell}+p_\mathrm{br}|_\mathrm{wall}$ consists of the sound-shell contribution~\eqref{eq:DetonaShell} and bubble-wall contribution~\eqref{eq:DetonaWall}, which we repeat here for your convenience,
\begin{align}
p_\mathrm{br}|_\mathrm{shell}&=-\frac{c_-^2}{1+c_-^2}(w_--w_s),\\
p_\mathrm{br}|_\mathrm{wall}&=-(w_+-w_-)+\frac{v_-}{v_--\xi_w}w_+.
\end{align}
It is straightforward to check that, by inserting the junction condition with $\bar{v}_+=\xi_w$ and $\bar{v}_{-}=\mu(\xi_w,v_-)$,
\begin{equation}
w_-=\frac{\bar{\gamma}_+^2\bar{v}_{+}}{\bar{\gamma}_-^2\bar{v}_-}w_+=\frac{\xi_w}{1-\xi_w^2}\frac{1-\mu(\xi_w,v_-)^2}{\mu(\xi_w,v_-)}w_+,
\end{equation}
the bubble-wall contribution to our hydrodynamic backreaction force exactly reproduces the pressure difference near the bubble wall,
\begin{equation}
p_\mathrm{br}|_\mathrm{wall}=\frac{v_-\xi_w}{1-v_-\xi_w}w_+= \overline{\Delta}(w\bar{\gamma}^2\bar{v}^2).
\end{equation}
As for proving the force balance, we can first rearrange the driving force~\eqref{eq:drnumodel} as
\begin{align}
p_\mathrm{dr}&=-\frac{c_-^2}{1+c_-^2}(w_--w_O)+\frac{c_-^2}{1+c_-^2}w_-\nonumber\\
&-\frac{c_{+}^2}{1+c_{+}^2}w_{N}+\frac{1}{1+c_{+}^2}\alpha_{+}w_{+}
\end{align}
to separate out the sound-shell contribution in the first term by noting $w_O=w(\xi=c_-)\equiv w_s$, and then the force balance would require the remaining terms to reproduce the wall contribution,
\begin{equation}
\frac{c_-^2}{1+c_-^2}w_--\frac{c_{+}^2}{1+c_{+}^2}w_{+}+\frac{1}{1+c_{+}^2}\alpha_{+}w_{+}=p_\mathrm{br}|_\mathrm{wall},
\end{equation}
leading to a relation
\begin{equation}
    \alpha_+=\frac{(1+c_+^2)[\xi_w+v_-(v_-\xi_wc_-^2-c_-^2-1)]}{(1+c_-^2)(v_--\xi_w)(v_-\xi_w-1)}-1,
\end{equation}
which is nothing but the plus-sign branch of \eqref{eq:vbarpm} with $\overline{v}_+=\xi_w$.

\subsection{Weak deflagration}
\label{subsec:ProveWeakDeflag}

For the weak deflagration mode, recall that our hydrodynamic evaluation of the full backreaction force $p_\mathrm{br}=p_\mathrm{br}|_\mathrm{wall}+p_\mathrm{br}|_\mathrm{shell}+p_\mathrm{br}|_\mathrm{shock}$ consists of the bubble-wall contribution~\eqref{eq:DeflagWall}, sound-shell contribution~\eqref{eq:DeflagShell}, and shock-wave-front contribution~\eqref{eq:DeflagShock}, which we repeat here for your convenience,
\begin{align}
p_\mathrm{br}|_\mathrm{wall}&=-(w_+-w_-)-\frac{v_+}{v_+-\xi_w}w_-,\\
p_\mathrm{br}|_\mathrm{shell}&=-\frac{c_+^2}{1+c_+^2}(w_L-w_+),\\
p_\mathrm{br}|_\mathrm{shock}&=-(w_R-w_L)+\frac{v_{sh}}{v_{sh}-\xi_{sh}}w_R.
\end{align}
It is straightforward to check that, by inserting the junction condition at the bubble-wall with $\bar{v}_+=\mu(\xi_w,v_+)$ and $\bar{v}_-=\xi_w$,
\begin{equation}
w_-=\frac{\bar{\gamma}_+^2\bar{v}_{+}}{\bar{\gamma}_-^2\bar{v}_-}w_+=\frac{\bar{v}_+}{1-\bar{v}_+^2}\frac{1-\xi_w^2}{\xi_w}w_+,
\end{equation}
the bubble-wall contribution to our hydrodynamic backreaction force exactly reproduces the pressure difference near the bubble-wall,
\begin{equation}
p_\mathrm{br}|_\mathrm{wall}=\frac{v_+^2-v_+\xi_w}{1-v_+^2}w_+= \overline{\Delta}(w\bar{\gamma}^2\bar{v}^2).
\end{equation}
Similarly, inserting the junction condition at the shock-wave front with $\tilde{v}_R=\xi_{sh}$, $\tilde{v}_L=\mu(\xi_{sh},v_{sh})$, and $\mu(\xi_{sh},v_{sh})\xi_{sh}=c_+^2$,
\begin{equation}
w_L=\frac{\tilde{\gamma}_{R}^2\tilde{v}_{R}}{\tilde{\gamma}_{L}^2\tilde{v}_{L}}w_R=\frac{\xi_{sh}}{1-\xi_{sh}^2}\frac{1-\mu(\xi_{sh},v_{sh})^2}{\mu(\xi_{sh},v_{sh})}w_R,
\end{equation}
also allows us to rewrite the shock-wave-front contribution as
\begin{equation}
p_\mathrm{br}|_\mathrm{shock}=\frac{c_+^2-\xi_{sh}^2}{\xi_{sh}^2-1}w_R=-\frac{c_+^2}{1+c_+^2}(w_R-w_L),
\end{equation}
so that we can further rearrange the driving force~\eqref{eq:drnumodel} as
\begin{align}
p_\mathrm{dr}&=-\frac{c_+^2}{1+c_+^2}[(w_N-w_L)+(w_L-w_+)]-\frac{c_+^2}{1+c_+^2}w_+\nonumber\\
&+\frac{c_-^2}{1+c_-^2}w_O+\frac{1}{1+c_+^2}\alpha_+ w_+
\end{align}
to separate out both the shock-wave-front and sound-shell contributions in the first term by noting $w_O=w_-$ and $w_N=w_R$; then the force balance would require the remaining terms to reproduce the bubble-wall contribution,
\begin{equation}
-\frac{c_+^2}{1+c_+^2}w_++\frac{c_-^2}{1+c_-^2}w_-+\frac{1}{1+c_{+}^2}\alpha_{+}w_{+}=p_\mathrm{br}|_\mathrm{wall},
\end{equation}
leading to a relation
\begin{align}
\alpha_+=&\frac{1}{(c_-^2+1)(v_+^2-1)\xi_w}\left[v_+\xi_w(\xi_w-v_+)+c_+^2\xi_w(v_+\xi_w-1)\right.\nonumber\\
&\left.+c_-^2(\xi_w-v_+-v_+c_+^2+v_+^2c_+^2\xi_w)\right],
\end{align}
which is nothing but the minus-sign branch of \eqref{eq:vbarpm} with $\bar{v}_-=\xi_w$.

\subsection{Jouguet detonation and deflagration}

For both the Jouguet detonation and Jouguet deflagration modes, recall that our hydrodynamic evaluation of the full backreaction force $p_\mathrm{br}=p_\mathrm{br}|_\mathrm{shell}^\mathrm{inner}+p_\mathrm{br}|_\mathrm{wall}+p_\mathrm{br}|_\mathrm{shell}^\mathrm{outer}+p_\mathrm{br}|_\mathrm{shock}$ consists of the bubble-wall contribution~\eqref{eq:JouguetWall}, shock-wave-front contribution~\eqref{eq:JouguetShock}, and the inner and outer sound-shell contributions~\eqref{eq:JouguetShell} inside and outside of the bubble wall, which we repeat here for your convenience,
\begin{align}
p_\mathrm{br}|_\mathrm{wall}&=-(w_+-w_-)+\frac{c_-(1-\xi_w^2)(v_--v_+)w_-}{(1-c_-^2)(v_--\xi_w)(\xi_w-v_+)},\\
p_\mathrm{br}|_\mathrm{shell}&=-\frac{c_-^2}{1+c_-^2}(w_--w_s)-\frac{c_+^2}{1+c_+^2}(w_L-w_+),\\
p_\mathrm{br}|_\mathrm{shock}&=-(w_R-w_L)+\frac{v_{sh}}{v_{sh}-\xi_{sh}}w_R.
\end{align}
It is straightforward to check that, by inserting the junction condition at the bubble-wall with $v_+=\mu(\xi_w,\overline{v}_+)$ and $v_-=\mu(\xi_w,c_-)$,
\begin{equation}
w_-=\frac{\bar{\gamma}_+^2\bar{v}_{+}}{\bar{\gamma}_-^2\bar{v}_-}w_+=\frac{\bar{v}_+}{1-\bar{v}_+^2}\frac{1-c_-^2}{c_-}w_+,
\end{equation}
the bubble-wall contribution to our hydrodynamic backreaction force exactly reproduces the pressure difference near the bubble wall,
\begin{equation}
p_\mathrm{br}|_\mathrm{wall}=\frac{(\bar{v}_+-c_-)\bar{v}_+}{1-\bar{v}_+^2}w_+=\overline{\Delta}(w\bar{\gamma}^2\bar{v}^2).
\end{equation}
Similarly, inserting the junction condition at the shock-wave front with $\tilde{v}_R=\xi_{sh}$, $\tilde{v}_L=\mu(\xi_{sh},v_{sh})$, and $\mu(\xi_{sh},v_{sh})\xi_{sh}=c_+^2$,
\begin{equation}
w_L=\frac{\tilde{\gamma}_{R}^2\tilde{v}_{R}}{\tilde{\gamma}_{L}^2\tilde{v}_{L}}w_R=\frac{\xi_{sh}}{1-\xi_{sh}^2}\frac{1-\mu(\xi_{sh},v_{sh})^2}{\mu(\xi_{sh},v_{sh})}w_R,
\end{equation}
also allows us to rewrite the shock-wave-front contribution as
\begin{align}
p_\mathrm{br}|_\mathrm{shock}=\frac{c_+^2-\xi_{sh}^2}{\xi_{sh}^2-1}w_R=-\frac{c_+^2}{1+c_+^2}(w_R-w_L),
\end{align}
so that we can further rearrange the driving force~\eqref{eq:drnumodel} as
\begin{align}
p_\mathrm{dr}=&-\frac{c_+^2}{1+c_+^2}(w_N-w_L)-\frac{c_+^2}{1+c_+^2}(w_L-w_+)\nonumber\\
&-\frac{c_-^2}{1+c_-^2}(w_--w_O)-\frac{c_+^2}{1+c_+^2}w_+\nonumber\\
&+\frac{c_-^2}{1+c_-^2}w_-+\frac{1}{1+c_+^2}\alpha_+ w_+
\end{align}
to separate out the shock-wave-front contribution and the outer and inner sound-shell contributions in the first three terms by noting  
$w_O=w_-$ and $w_N=w_R$, and then the force balance would require the remaining terms to reproduce the bubble-wall contribution,
\begin{equation}
-\frac{c_+^2}{1+c_+^2}w_++\frac{c_-^2}{1+c_-^2}w_-+\frac{1}{1+c_+^2}\alpha_+ w_+=p_\mathrm{br}|_\mathrm{wall},
\end{equation}
leading to a relation
\begin{equation}
\alpha_+=\frac{[c_+^2(1+c_-^2-2c_-\bar{v}_+)+\bar{v}_+(\bar{v}_+-2c_-+\bar{v}_+c_-^2)]}{(1+c_-^2)(1-\bar{v}_+^2)},
\end{equation}
which is nothing but the plus-sign and minus-sign branches of \eqref{eq:vbarpm} with $\bar{v}_-=c_-$ for Jouguet detonation and Jouguet deflagration modes, respectively.

\section{Conclusions}\label{sec:conclusion}

\begin{figure*}
\centering
\includegraphics[width=\textwidth]{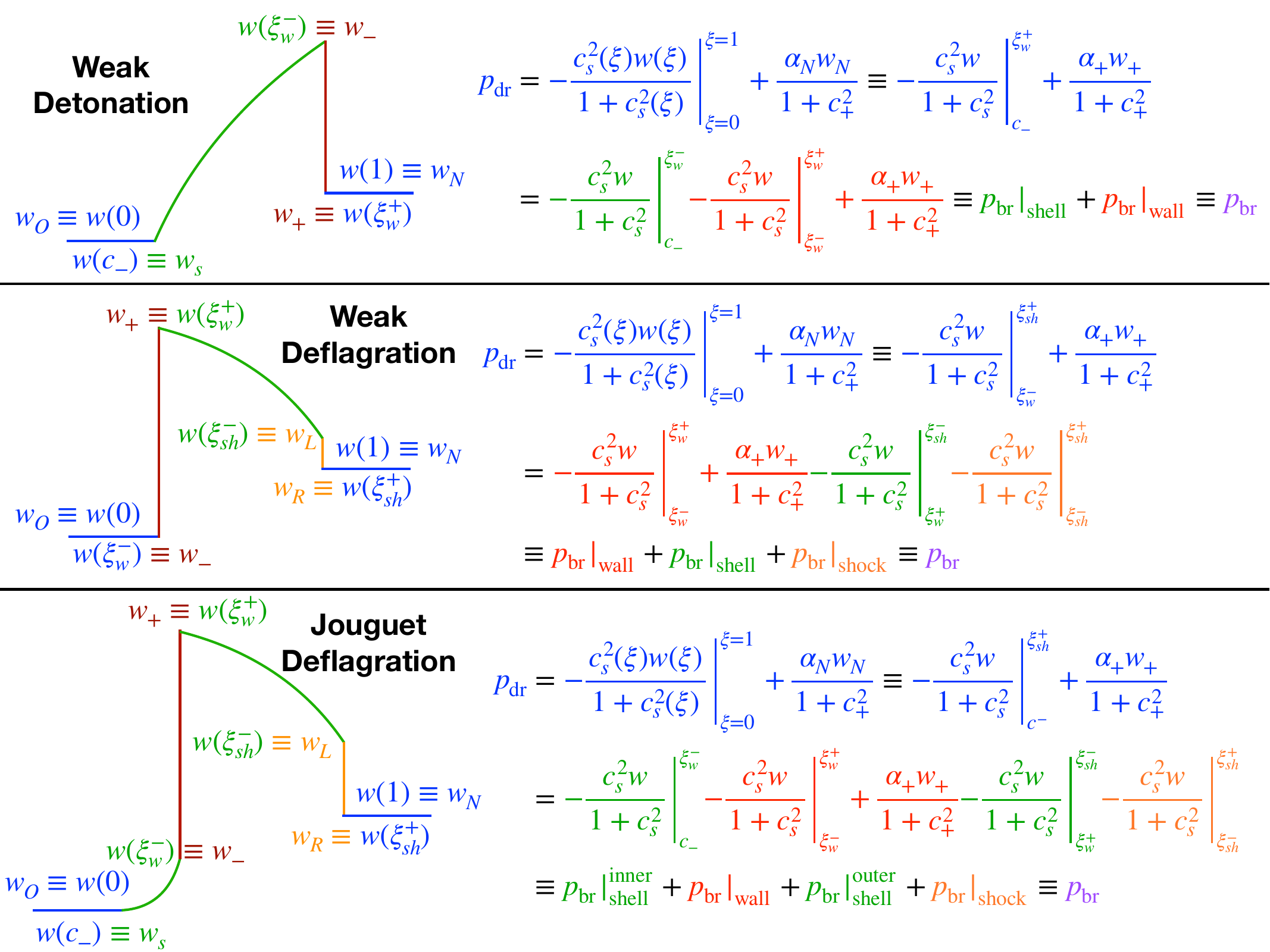}\\
\caption{A concise summary of our hydrodynamic backreaction force (along with all its contributions highlighted with corresponding colors) beyond the bag EOS in the $\nu$-model case for bubble expansions of (top) weak detonation, (middle) weak deflagration, and (bottom) Jouguet deflagration modes (the Jouguet detonation case is similar to the Jouguet deflagration case and hence not shown here for clarity). The sound velocity is defined by a steplike function as $c_s(\xi)=c_-\Theta(\xi_w-\xi)+c_+\Theta(\xi-\xi_w)$.}\label{fig:BackReactionForce}
\end{figure*}

The cosmological FOPT proceeds via stochastic bubble nucleations followed by a rapidly accelerating expansion until approaching an asymptotic expansion before bubble percolations of colliding walls. The asymptotic expansion stage can be well approximated as a nonrunaway steady-state expansion of a thin wall in a thermal plasma that can be well described as a perfect fluid along with the expanding scalar wall. The nearly constant expansion of the scalar wall is maintained via an exact balance between the driving force and the backreaction force including not only the friction force from the nonequilibrium effect but also the thermal force from the inhomogeneous temperature profile across the discontinuous interfaces. 

Although both the thermal force and friction force cannot be known exactly without extra inputs from the junction conditions other than those from a conserved energy-momentum tensor, their summation as the total backreaction force can be explicitly computed purely from the perfect-fluid hydrodynamics alone without assuming thermal equilibrium across the discontinuous interfaces. This hydrodynamic evaluation of the total backreaction force has been confirmed in our previous study both numerically and analytically but only for a bag EOS. 

In this paper, we have successfully applied our hydrodynamic evaluation of the total backreaction force to the case beyond the bag model with a $\nu$-model EOS. The hydrodynamic backreaction force in this case has also been analytically checked to not only exactly balance the driving force but also exactly reproduce the pressure difference near the wall from the junction condition at the bubble wall. The results are summarized in Fig.~\ref{fig:BackReactionForce} for the bubble expansion of weak detonation, weak deflagration, and Jouguet deflagration types (the similar case with Jouguet detonation is not shown here for clarity).

It is evident from Fig.~\ref{fig:BackReactionForce} that the final expression for calculating each contribution (wall, shell, and shock, if any) to the total backreaction force boils down to evaluate the corresponding difference of the enthalpy (up to a constant factor from the sound velocity profile, which, however, is not necessarily continuous across the wall as expected from a general EOS) ,
\begin{align}
-\frac{c_s^2(\xi)w(\xi)}{1+c_s^2(\xi)},
\end{align}
except for an extra term $\alpha_+w_+/(1+c_+^2)$ in the case of the wall contribution to the total backreaction force. Apart from the spatial decomposition into wall, shell, and shock contributions, the total backreaction force can also be decomposed into the thermal and friction forces according to their properties~\eqref{eq:balance}, where the out-of-equilibrium effect only contributes to the friction force that can only be nonzero at the discontinuous interface such as the bubble wall or shock front, if any, as shown shortly below in Eq.~\eqref{eq:shellfriction}. 

It is worth noting that all conclusions obtained in this paper rely on the thin-wall assumption as imposed in the third paragraph of Sec.~\ref{subsec:wallfluid}, resulting in the discontinuous jumping for various quantities across the wall. It is this discontinuous jumping that contributes nontrivially to the backreaction force from integrating one discontinuous quantity over another discontinuous quantity, which can be rigorously evaluated using a mathematical trick introduced in our previous paper~\cite{Wang:2022txy} as also noted below Eq.~\eqref{eq:difftilde}. When computing the thermal and friction forces, the nontrivial contributions involve integrations such as $\int s\mathrm{d}T$ and $\int T\mathrm{d}s$, respectively, which would require some continuous functions $s(T)$ and $T(s)$ relating the discontinuous profiles of the temperature $T(\xi)$ and entropy density $s(\xi)$ across the discontinuous interface that can only be provided by some extra junction conditions other than the one from the conserved energy-momentum tensor. Fortunately, when computing the thermal and friction forces combined, namely the total backreaction force, this nontrivial contribution involves $\int w\mathrm{d}v$, which is fully determined by a continuous function $w(v)$ that can be simply obtained by the junction condition from the usual conservation of the energy-momentum tensor. Therefore, the local equilibrium assumption $p_{\delta f}=0$ that eliminates the friction force has never been used when computing the total backreaction force.

In short summary, our hydrodynamic evaluation of the total backreaction force does rely on the thin-wall approximation but not necessarily the local equilibrium assumption, without which the enthalpy profile for computing the backreaction force can still be determined by hydrodynamics alone as long as the steady-state expansion is reached. Future investigations should be implemented from going beyond the thin-wall approximation (see, for example,  a recent study~\cite{Ai:2023suz}) and the steady-state assumption (for example, during the acceleration stage). All this progress would be fully appreciated for the final goal to determine the terminal wall velocity.

\section{Discussions}\label{sec:discussions}

Before we move to the discussion part, we would like to clarify the significance of defining such a backreaction force~\eqref{eq:balance}:

First of all, the traditional approach~\cite{Bodeker:2009qy,Bodeker:2017cim,BarrosoMancha:2020fay,Hoche:2020ysm,Azatov:2020ufh,Gouttenoire:2021kjv,DeCurtis:2022hlx} to compute the terminal wall velocity is to first calculate the pressure recoil (usually some function of the wall velocity) from interacting particles acting on the bubble wall alone in the ultrarelativistic limit, and then simply solve the wall velocity from balancing the driving force with this wall pressure recoil. However, this traditional approach is incomplete since it is actually the backreaction force $p_\mathrm{br}=p_\mathrm{th}+p_\mathrm{fr}=p_\mathrm{br}|_\mathrm{wall}+p_\mathrm{br}|_\mathrm{shell}+p_\mathrm{br}|_\mathrm{shock}(\mathrm{if\,\,any})$ instead of the wall pressure difference $p_\mathrm{br}|_\mathrm{wall}=p_--p_+$ alone that eventually balances the driving force $p_\mathrm{dr}=\Delta V_\mathrm{eff}$ unless the effective potential difference is not taken between the innermost and outermost ends of the bubble profile but in the vicinity near the wall.\footnote{Note that the traditional approach of force balance is equivalent to ours as one can explicitly prove that $p_\mathrm{dr}|_\mathrm{wall}=p_\mathrm{th}|_\mathrm{wall}+p_\mathrm{fr}|_\mathrm{wall}=p_\mathrm{br}|_\mathrm{wall}$, where the difficultly of computing $p_\mathrm{br}|_\mathrm{shell}+p_\mathrm{br}|_\mathrm{shock}$ has been moved to compute $p_\mathrm{dr}|_\mathrm{wall}\equiv\overline{\Delta}V_\mathrm{eff}\equiv V_\mathrm{eff}(\phi_+,T_+)-V_\mathrm{eff}(\phi_-,T_-)$ with the near-wall temperatures $T_+\equiv T(\xi_w^+)$ and $T_-\equiv T(\xi_w^-)$ separated from the far-end temperatures $T_N\equiv T(\xi=1)$ and $T_O\equiv T(\xi=0)$ by the sound-shell and shock-front (if any) parts of fluid motions.}  Even if we manage to acquire full knowledge of the traditional wall pressure recoil, there are still the contributions from the sound-shell and shock-front (if any) parts to be determined so as to compute the terminal wall velocity.  In fact, even for the detonation expansion of an ultrarelativistic wall, the pressure difference from the sound-shell is not negligible as one might naively anticipate when eventually balancing the driving force together with the wall pressure difference. This can easily be seen from our previous studies~\cite{Wang:2022txy,Li:2023xto} that the total backreaction force numerically approaches $p_\mathrm{br}\to-\frac14\frac{(4/5)\alpha_Nw_N}{(2/3)+\alpha_N}+\frac34\alpha_Nw_N$, while the wall pressure difference analytically approaches $p_\mathrm{br}|_\mathrm{wall}\to\frac32\alpha_Nw_N$, rendering the remaining shell contribution to the backreaction force asymptotic to $p_\mathrm{br}|_\mathrm{shell}\to-\frac14\frac{(4/5)\alpha_Nw_N}{(2/3)+\alpha_N}-\frac34\alpha_Nw_N$, whose absolute value is at least half of the wall pressure difference, $|p_\mathrm{br}|_\mathrm{shell}|/p_\mathrm{br}|_\mathrm{wall}\to\frac12+\frac{2}{5(2+3\alpha_N)}>\frac12$. Therefore, the introduction of the backreaction force helps to complete the traditional strategy to determine the terminal wall velocity. 

Next, the traditional approach by calculating the wall pressure recoil from solving the Boltzmann equations only accounts for the friction force $p_\mathrm{fr}$ from the nonequilibrium effects but overlooks the thermal force $p_\mathrm{th}$ from the inhomogeneous temperature profile across the bubble wall and sound-shell as well as shock front (if any). This thermal force was previously absorbed into a redefinition of the driving force in early studies~\cite{Espinosa:2010hh,Konstandin:2010dm} but with its wall contribution $p_\mathrm{th}|_\mathrm{wall}$ naively estimated by the averaged value across the wall assuming a homogeneous temperature profile. As we will show shortly below in Appendix~\ref{app:FthFfr}, the wall contributions of both thermal force and friction force require an extra input from a continuous function $T(s)$ relating the discontinuous temperature and entropy density profiles across the bubble wall, and only the combination of the thermal force and friction force into the backreaction force can render their nontrivial wall contributions into a form that can be exactly evaluated from hydrodynamics with a given EOS.

Last but not least, the introduction of backreaction force clarifies a previous misleading claim~\cite{BarrosoMancha:2020fay,Balaji:2020yrx} that the pressure difference $\overline{\Delta}(\bar{\gamma}^2\bar{v}^2w)=(\gamma_w^2-1)\overline{\Delta}w$ against the wall expansion might be proportional to the factor $(\gamma_w^2-1)$ involving the Lorentz factor $\gamma_w\equiv(1-\xi_w)^{-1/2}$ of the terminal wall velocity $\xi_w$. In fact, as shown in our previous study~\cite{Wang:2022txy} of the backreaction force, the difference $\overline{\Delta}(\bar{\gamma}^2\bar{v}^2w)$ taken near the wall does not equal $(\gamma_w^2-1)\overline{\Delta}w$ but exactly reproduces the wall contribution of backreaction force, $\overline{\Delta}(\bar{\gamma}^2\bar{v}^2w)=p_\mathrm{br}|_\mathrm{wall}=p_--p_+$, namely the opposite of wall pressure difference $\Delta p\equiv p_+-p_-$ as expected from the junction condition~\eqref{eq:Wall2ndJunction}. Although taking the difference away from the wall for $\Delta(\bar{\gamma}^2\bar{v}^2w)=(\gamma_w^2-1)\Delta w$ indeed produces the factor $(\gamma_w^2-1)$,  it is unfortunately not the pressure difference $-\Delta p\equiv p_O-p_N\neq\Delta(\bar{\gamma}^2\bar{v}^2w)$ that actually balances the driving force since $-\overline{\Delta}p=\overline{\Delta}(\bar{\gamma}^2\bar{v}^2w)$ is only valid when the difference is taken near the wall instead of taking the difference away from the wall, $-\Delta p\neq\Delta(\bar{\gamma}^2\bar{v}^2w)$. Therefore, one cannot conclude that the pressure difference against the wall expansion is proportional to $(\gamma_w^2-1)$.

To eventually reveal the $\gamma_w$-scaling behavior for the backreaction force $p_\mathrm{br}(\xi_w)$ as a function of the terminal wall velocity $\xi_w$, the current steady-state hydrodynamic approximation is not enough since the steady-state expansion is only reached when the backreaction force has already balanced the driving force $p_\mathrm{dr}=\Delta V_\mathrm{eff}$, which itself is a constant given by $\Delta V_\mathrm{eff}$ independent of $\xi_w$. We will go beyond the steady-state hydrodynamic approximation to extract the function $p_\mathrm{br}(\xi_w)$ during the accelerating stage of bubble expansion in a series of future works step by step. Several discussions are given below regarding this final goal.

\subsection{Most general EOS with varying sound velocity}\label{app:GeneralEoS}

The sound velocity defined as $c_s^2=\partial_\xi p/\partial_\xi e$ is a function of the temperature, which itself is a function of the self-similar coordinate $\xi$. Therefore, the most general EOS beyond the $\nu$ model should also admit a varying profile for the sound velocity $c_s(\xi)$ beyond the simple steplike ansatz $c_s(\xi)=c_-\Theta(\xi_w-\xi)+c_+\Theta(\xi-\xi_w)$. One such example is given in our previous study~\cite{Wang:2022lyd}. In general, all beyond-bag-model EOS can be formally parametrized as
\begin{align}
p(T)=\frac13 a(T)T^4-\epsilon(T), \quad  e(T)=a(T)T^4+\epsilon(T)
\end{align}
in terms of two baglike quantities
\begin{align}
a(T)=\frac{3}{4T^3}\frac{\partial p}{\partial T}=\frac{3w}{4T^4},\quad \epsilon(T)=\frac{e(T)-3p(T)}{4}
\end{align}
with a constraint relation
\begin{align}
\frac{\partial\epsilon}{\partial T}=\frac{T^4}{3}\frac{\partial a}{\partial T}.
\end{align}
Therefore, the sound velocity deviates from the bag model by the temperature dependence in the derivative of $a(T)$ or $\epsilon(T)$ as
\begin{align}\label{eq:csT}
c_s^2(T(\xi))=\frac{\partial_T p}{\partial_T e}=\frac13\frac{aT^3}{aT^3+\frac{\partial\epsilon}{\partial T}}=\frac13\frac1{1+\frac13\frac{\partial\ln a}{\partial\ln T}}.
\end{align}
Explicitly calculating the backreaction force in this case along with its decompositions within the sound-shell and across discontinuous interfaces should be challenging. Nevertheless, it is inspiring to look at the analytic results from the $\nu$-model EOS in Fig.~\ref{fig:BackReactionForce} that the final results for the most general EOS might just stay the same as long as the varying profile of the sound velocity $c_s(\xi)$ is used. We will leave this hydrodynamic backreaction force in the most general EOS for future works.

\subsection{Hydrodynamic evaluation of thermal force and friction force}\label{app:FthFfr}

As we have shown in Sec.~\ref{sec:setup}, the total backreaction force can be decomposed in two ways: one is to decompose it according to its components into the thermal force and friction force, and the other is to decompose it according to its contributions into the sound-shell and discontinuous-interface parts. The total backreaction force~\eqref{eq:balance} admits a hydrodynamic expression~\eqref{eq:brhydro}, which can be further evaluated using the perfect-fluid hydrodynamics alone with its discontinuous contribution evaluated by the junction conditions from the conserved energy-momentum tensor across the bubble wall and shock-wave front (if any). It is tempting to ask whether the thermal force and friction force could also exhibit hydrodynamic expressions and can be further evaluated individually using perfect-fluid hydrodynamics alone. A short answer to the former one (hydrodynamic expression) is yes but the latter one (hydrodynamic evaluation) is no, which we will elaborate briefly below.

The hydrodynamic expression for the thermal force is straightforward by adopting the most general EOS prementioned,
\begin{align}\label{eq:themo}
p_\mathrm{th}
&\equiv\int\mathrm{d}\xi\frac{\mathrm{d}T}{\mathrm{d}\xi}\frac{\partial V_\mathrm{eff}}{\partial T}=-\int\mathrm{d}\xi\frac{\mathrm{d}T}{\mathrm{d}\xi}\frac{\partial p}{\partial T}\nonumber\\
&=\int\mathrm{d}\xi\frac{\mathrm{d}T}{\mathrm{d}\xi}\left(\frac{\partial\epsilon}{\partial T}-\frac13\frac{\partial a}{\partial T}T^4-\frac43 a(T) T^4\right)\nonumber\\
&=\int\mathrm{d}\xi\frac{\mathrm{d}T}{\mathrm{d}\xi}\left(-\frac43a(T)T^3\right)\equiv-\int s(T)\mathrm{d}T,
\end{align}
where $s(T)\equiv w(T)/T=[e(T)+p(T)]/T=(4/3)a(T)T^3$ is the entropy density. Then, it is easy to evaluate the sound-shell part of the thermal force using perfect-fluid hydrodynamics alone. For example, for a bag EOS, we can explicitly write it down as
\begin{align}
p_\mathrm{th}|_\mathrm{shell}&=-\frac13\int_\mathrm{shell}\mathrm{d}(aT^4)=-\frac14\Delta_\mathrm{shell}w\nonumber\\
&=\begin{cases}
-\frac14(w_--w_s),&\quad \text{detonation},\\
-\frac14(w_L-w_+),&\quad \text{deflagration},\\
-\frac14(w_L-w_++w_--w_s),&\quad \text{hybrid}.
\end{cases}
\end{align}
However, the discontinuous part of the thermal force involves integration over $s\mathrm{d}T$ in the vicinity of a discontinuous interface, where both $s$ and $T$ experience a sudden jump across the discontinuous interface. This kind of integration can be rigorously evaluated using a mathematical trick we introduced in Ref.~\cite{Wang:2022txy} as long as a continuous function $s(T)$ could be provided as an extra junction condition across the discontinuous interface. Unfortunately, the conserved energy-momentum tensor only provides a continuous function $w(v)$ from the junction condition across the discontinuous interface.

Based on the same reason, the friction force also admits a hydrodynamic expression but its discontinuous part cannot be evaluated using hydrodynamics alone. To see this, we first expand the left-hand side of~\eqref{eq:EnthalpyFlow} with $\nabla_\mu(wu^\mu)=T\nabla_\mu(su^\mu)+(su^\mu)\nabla_\mu T$, and then rewrite the first term of the right-hand side of~\eqref{eq:EnthalpyFlow} with $u^\mu\nabla_\mu T(\partial V_\mathrm{eff}/\partial T)=-u^\mu\nabla_\mu T(\partial p/\partial T)=-u^\mu(\nabla_\mu T)s$; thus, we can convert the conservation violation equation of enthalpy flow~\eqref{eq:EnthalpyFlow} into a conservation violation equation of entropy flow as
\begin{align}
T\nabla_\mu(su^\mu)=-u^\mu\nabla_\mu\phi\frac{\partial p_{\delta f}}{\partial\phi}.
\end{align}
Therefore, we can similarly obtain the hydrodynamic expression for the friction force,
\begin{align}\label{eq:pfrhydro}
p_\mathrm{fr}
&=\int\mathrm{d}\xi\frac{\mathrm{d}\phi}{\mathrm{d}\xi}\frac{\partial p_{\delta f}}{\partial\phi}=\int\mathrm{d}\xi\frac{Tt\nabla_\mu(su^\mu)}{\gamma(\xi-v)}\nonumber\\
&=\int_0^1\mathrm{d}\xi\left(-T\frac{\mathrm{d}s}{\mathrm{d}\xi}+\frac{2wv}{\xi(\xi-v)}+\frac{w\gamma^2}{\mu(\xi,v)}\frac{\mathrm{d}v}{\mathrm{d}\xi}\right).
\end{align}
It is straightforward to compute the sound-shell part of the friction force using the hydrodynamic EOMs~\eqref{eq:dv} and~\eqref{eq:dw} in replacement of $v/\xi$ and $\mathrm{d}v/\mathrm{d}\xi$ terms as
\begin{align}\label{eq:shellfriction}
p_\mathrm{fr}|_\mathrm{shell}&=\int\frac{\mathrm{d}w}{1+c_s^2}-T\mathrm{d}s\nonumber\\
&=\int\mathrm{d}T\left[\frac{4a(T)T^3}{3(1+c_s^2)}-\frac{4c_s^2a(T)T^3}{1+c_s^2}-\frac{4c_s^2a'(T)T^4}{3(1+c_s^2)}\right]\nonumber\\
&=0,
\end{align}
where the last two equalities are reached even for the most general EOS~\eqref{eq:csT}. Therefore, the friction force receives contribution only from the discontinuous-interface part. Once again, the integral over $T\mathrm{d}s$ across a discontinuous interface would also require an extra input from a continuous function $T(s)$ relating discontinuous profiles $T$ and $s$ in the vicinity of the discontinuous interface, which cannot be provided by the junction conditions from the conserved energy-momentum tensor.

Nevertheless, when combing the thermal force and friction force into the total backreaction force, the $s\mathrm{d}T$ and $T\mathrm{d}s$ terms are added up to a total derivative $\mathrm{d}w$, which can now be computed by the hydrodynamics alone. This is the main reason that we can contribute the thermal force to the total backreaction force instead of a modification of the driving force as previously thought~\cite{Espinosa:2010hh,Konstandin:2010dm}. We will evaluate in a future study both the thermal force and friction individually by proposing such a junction condition relating the temperature and entropy density across the discontinuous interface from microscopic particle physics.

\subsection{General bubble expansion at strong coupling}\label{app:GeneralStrongCoupling}

One possible application of our hydrodynamic evaluation of the total backreaction force is to estimate the phase pressure difference $p_O-p_N$ away from the wall in the nonrelativistic limit of terminal wall velocity for a strongly coupled FOPT with a bag EOS~\cite{Li:2023xto}, reproducing an intriguing linear correlation for a planar wall that was first observed in holographic numerical simulations,
\begin{align}
p_O-p_N=\frac{\alpha_N w_N}{c_s}\xi_w+\mathcal{O}(\xi_w^2).
\end{align}
We also predict a quadratic-logarithmic relation for a cylindrical wall, and in particular, a quadratic relation for a spherical wall,
\begin{align}
p_O-p_N=\alpha_N w_N\left(\frac{1}{c_s^2}+\frac{\alpha_N}{c_s^4+c_s^6}\right)\xi_w^2+\mathcal{O}(\xi_w^4),
\end{align}
reversing which allows us to estimate the terminal wall velocity from hydrodynamics by 
\begin{align}
\xi_w&=\sqrt{\frac{c_s^4\delta}{\alpha_N+c_s^2+c_s^4}},\\
\delta=\frac{\Delta V_\mathrm{eff}}{\Delta V_0}&=\left(1-\frac{c_s^2}{1+c_s^2}\frac{\partial\ln\Delta V_\mathrm{eff}}{\partial\ln T}\right)^{-1}.
\end{align}
On the other hand, the pressure difference near the wall is found to behave universally as
\begin{align}
p_+-p_-=\alpha_N w_N\left(\frac{1}{c_s^2}-\frac{\alpha_N}{c_s^4}\right)\xi_w^2+\mathcal{O}(\xi_w^3)
\end{align}
regardless of the wall geometries. In a following paper~\cite{Wang:2023lam}, we have generalized these nonrelativistic behaviors beyond the bag EOS, which can be explicitly tested against future holographic numerical simulations of strongly coupled FOPTs.

\subsection{Hydrodynamics during accelerating expansion stage}\label{app:HydroAccele}

As the solely undetermined phenomenological parameter in characterizing the energy-density spectra of stochastic gravitational wave backgrounds from FOPTs, the terminal wall velocity cannot be naively estimated from the friction force alone since the thermal force also contributes to the total backreaction force acting on the bubble wall, and more importantly, it is the total backreaction force consisting of both sound-shell and bubble-wall as well as shock front (if any) parts that gradually balance the driving force to eventually approach the terminal wall velocity. Therefore, the final determination of the terminal wall velocity would require a full understanding of the total backreaction force for the bubble expansion during both accelerating expansion and asymptotic expansion stages. 

For the asymptotic expansion stage, it is the late stage of FOPT that is under considered so that the approximation for a thin-wall steady-state self-similar expansion can be used, which naturally induces the perfect-fluid hydrodynamics. However, for the accelerating expansion stage, no such privileges can be made, and the integrated scalar EOM,
\begin{align}
\int\mathrm{d}\bar{r}\frac{\mathrm{d}\phi}{\mathrm{d}\bar{r}}\left(\nabla^2\phi-\frac{\partial V_\mathrm{eff}}{\partial\phi}+\frac{\partial p_{\delta f}}{\partial\phi}\right)=0,
\end{align}
seems to induce an effective EOM for a scalar-wall expansion (see Sec. 3 of Ref.~\cite{Cai:2020djd}),
\begin{align}
\sigma\gamma_w^3\ddot{r}_w+\frac{2\sigma\gamma_w}{r_w}=\Delta V_\mathrm{eff}-\int\mathrm{d}\bar{r}\frac{\mathrm{d}T}{\mathrm{d}\bar{r}}\frac{\partial V_\mathrm{eff}}{\partial T}+\int\mathrm{d}\bar{r}\frac{\mathrm{d}\phi}{\mathrm{d}\bar{r}}\frac{\partial p_{\delta f}}{\partial\phi},
\end{align}
where $\bar{r}=\gamma_w(t)[r-r_w(t)]$ and $\bar{t}=\gamma_w(t)[t-v_w(t)r]$ are coordinates after a Lorentz boost with the bubble-wall velocity $v_w(t)=\dot{r}_w(t)$ and corresponding Lorentz factor $\gamma_w(t)=1/\sqrt{1-v_w^2(t)}$, and $\sigma\equiv\int_{-\infty}^{+\infty}\phi'(\bar{r})^2\mathrm{d}\bar{r}$ is the bubble tension. It is unclear how the hydrodynamics looks so as to compute the last two integrals. It is also questionable to arrive at the left-hand side since $\bar{r}$ can only reach $\pm\infty$ at a late-time limit with $r_w(t\to\infty)\to\infty$ for $r=0$. We will derive an alternative effective EOM for the scalar wall in a upcoming work.

\begin{acknowledgments}
We thank Wen-Yuan Ai for the insightful discussions, as well as an anonymous referee.
S.J.W. is supported by the National Key Research and Development Program of China Grants No. 2021YFC2203004, No. 2020YFC2201501, and No. 2021YFA0718304; 
the National Natural Science Foundation of China Grants 
%No. 11647601, No. 11821505, No. 11851302, No. 12047503, No. 11991052, No. 12075297, No. 12047558, and 
No. 12105344, No. 12235019, and No. 12047503;
the Key Research Program of the Chinese Academy of Sciences (CAS) Grant No. XDPB15; 
the Key Research Program of Frontier Sciences of CAS; 
and the Science Research Grants from the China Manned Space Project with No. CMS-CSST-2021-B01.
\end{acknowledgments}

\bibliography{ref}

\end{document}